\documentclass[aps,prx,twocolumn,superscriptaddress,nofootinbib,notitlepage,longbibliography]{revtex4-1}

\usepackage{amsbsy}
\usepackage{amsfonts}
\usepackage{amsmath}
\usepackage{amstext}
\usepackage{amsthm}
\usepackage{amssymb}
\usepackage{bm}
\usepackage{color}
\usepackage{comment}
\usepackage{booktabs}
\usepackage{datetime}
\usepackage{dsfont}
\usepackage{dcolumn}
\usepackage{esint}
\usepackage{extarrows}
\usepackage{graphicx}
\usepackage[colorlinks,citecolor=blue]{hyperref} \usepackage{cleveref}
\usepackage{longtable,booktabs}
\usepackage{mathrsfs}
\usepackage{multirow}
\usepackage[super]{nth}
\usepackage{subfigure}
\usepackage{tensor}
\usepackage{url}
\usepackage{xfrac}

\hypersetup{colorlinks=true, linkcolor=blue, filecolor=magenta, urlcolor=blue,}

\begin{document}
\graphicspath{{figures/}}


\title{Continuous transition from a Landau quasiparticle to a neutral spinon}  

\author{Jing-Yu Zhao} \affiliation{Institute for Advanced Study, Tsinghua University, Beijing 100084, China} 
\author{Shuai A. Chen} \affiliation{Department of Physics, Hong Kong University of Science and Technology, Clear Water Bay, Hong Kong 999077, China}
\author{Rong-Yang Sun} \affiliation{Computational Materials Science Research Team, RIKEN Center for Computational Science (R-CCS), Kobe, Hyogo 650-0047, Japan}
\affiliation{Quantum Computational Science Research Team, RIKEN Center for Quantum Computing (RQC), Wako, Saitama 351-0198, Japan}
\author{Zheng-Yu Weng} \affiliation{Institute for Advanced Study, Tsinghua University, Beijing 100084, China}




\date{\today}

\begin{abstract} 
We examine a wavefunction ansatz in which a doped hole can experience a quantum transition from a charge $+e$ Landau quasiparticle to a neutral spinon as a function of the underlying spin-spin correlation. As shown variationally, such a wavefunction accurately captures all the essential features revealed by exact diagonalization and density matrix renormalization group simulations in a two-leg $t$-$J$ ladder. Hence its analytic form can provide an explicit understanding of the mechanism for the unconventional ground state. The transition in the phase diagram is accompanied by a change of the hole composite from a tight charge-spin binding to a loosely-bound hole-spin pair. In the latter, the hole carries a \emph{finite} spin current but with vanishing charge current in the degenerate ground states.  We show that the charge of the hole composite here is dynamically diminished due to an internal relative hole-spin motion, which is fundamentally distinct from a simple charge-spin separation in a one-dimensional case. We further show that the same effect is also responsible for a strong pairing between two doped holes in such a non-Landau quasiparticle regime.     
\end{abstract}

\maketitle

\tableofcontents

\section{Introduction}\label{sec:intro}

How to properly characterize a single-particle excitation is one of the most essential challenges 
in the study of the strongly correlated Mott insulators \cite{Mott1949,Anderson1987,Imada1998,Lee2006}. 
In particular, a single chargon (hole) may serve as a building block for constructing a doped Mott insulator. The central issue under debate is whether such a chargon moving a quantum spin background will still behave like a conventional Landau-type quasiparticle or be fundamentally renormalized into a non-Landau quasiparticle via ``twisting'' the surrounding many-body spins in the background. 

In analogy to the electronic ``cloud'' associated with a Landau quasiparticle, it was widely believed earlier on that such a hole in the Mott insulator may still be the Landau-type after considering the longitudinal spin-polaron effect \cite{Schmitt-Rink1988,Kane1989}, which involves a distortion in the amplitude of the local spin magnetization around the hole. On the other hand, it was conjectured \cite{Anderson1990} that a nontrivial many-body response from the Mott insulator background may lead to an ``unrenormalizable Fermi-surface phase shift'', which can result in an ``orthogonality catastrophe'' to turn the doped hole into a non-Landau quasiparticle. 

Such an ``unrenormalizable phase shift'' has been explicitly identified with the phase-string effect in the $t$-$J$ \cite{Sheng1996,Wu2008} and Hubbard \cite{Zhang2014} models as a singular nonintegrable Berry phase acquired by the doped hole(s) moving in the quantum spin background of the Mott insulator. Physically, it implies that the hopping of the doped hole should generate a spin-current backflow. Exact diagonalization (ED) and density matrix renormalization group (DMRG) simulations have recently confirmed \cite{Zheng2018b} such an unconventional behavior of the doped hole in the two-dimensional (2D) square lattice. The hidden spin-current backflow overlooked by previous numerical works has been revealed \cite{Zheng2018b} to accompany and facilitate the hopping of the hole, leading to a nontrivial quantum number as a direct manifestation of a non-Landau-like quasiparticle. 
Making use of the variational Monte Carlo (VMC) method, a single-hole wavefunction ansatz has been constructed \cite{Chen2019}, which well interprets the numerical results including the nontrivial angular momenta $L_z=\pm 1$ under a $C_4$ rotational symmetry for a finite-size 2D sample up to $8\times 8$ \cite{Zheng2018b}. 

The phase-string effect or the spin-current backflow here leads to a \emph{transverse} spin twist to renormalize the doped hole. But previously a Landau quasiparticle behavior had been still inferred in a semiclassical field-theory approach \cite{Shraiman1988a,Shraiman1989} even with incorporating a long-range transverse dipolar spin twist beyond the longitudinal spin-polaron effect. It is because a singular coupling between the doped hole and the spin currents at the \emph{short distance} was omitted \cite{Weng1991}. The latter effect has been carefully and consistently implemented in the wavefunction approach \cite{Chen2019} to reproduce the correct behavior. Nevertheless, in spite of the good agreement of the VMC result \cite{Chen2019} and exact numerics \cite{Zheng2018b} at finite (small) sizes, the 2D single-hole problem can be further complicated by the antiferromagnetic (AF) long-range order which sets in as the thermodynamic limit is taken, which may further lead to a self-localization of the hole.  Thus, a thorough understanding of the 2D single-hole problem has to handle both singular effects in short-range and long-range physics together, which is beyond the finite-size exact numerical methods.  

On the other hand, it would be interesting to first focus on the singular short-range physics, which should be generally present even though the long-range AF order is expected to disappear at finite doping. Besides the above VMC study on a single hole in a finite-size system in 2D,  a single-hole-doped $t$-$J$ model on a two-leg ladder may be a more suitable ``toy'' system to examine such an effect as here the spin-spin correlation length remains finite even at half-filling \cite{Zhu2013}. The anisotropy of a two-leg ladder system can also provide us more tools to continuously tune the correlation of the spin background. The previous DMRG studies \cite{Zhu2013,Zhu2015b,Zhu2015} have established an exotic ground state for the single hole, which behaves like a non-Landau-quasiparticle with breaking charge translational symmetry \cite{Zhu2018a}. Note that the above non-Landau-quasiparticle picture was previously contested by another DMRG study \cite{White2015}, in which a finite quasiparticle spectral weight $Z_k$ has been explicitly measured. To reconcile both DMRG results, it was pointed out \cite{Zhu2018a} that the Landau's one-to-one correspondence principle is actually broken down even though $Z_k$ still remains finite in the non-Landau-quasiparticle regime, where the hole object is of a two-component structure, composed of a translation-invariant Bloch wave component and a charge-incoherent component. In other words, the DMRG results indicate the existence of a new type of non-Landau quasiparticle. Especially a quantum transition for the doped hole to become a true Landau quasiparticle has been also found by reducing the spin-spin correlation length along the ladder direction \cite{Zhu2015b,Zhu2015,Zhu2018a,White2015}, with the translation symmetry and charge coherence being eventually restored. Therefore, a wavefunction description of such a single-hole problem in a two-leg ladder can offer a valuable proof-of-principle on how a bare doped hole can be specifically ``twisted'' into a non-Landau-quasiparticle due to the intrinsic Mott physics  even in the absence of an AF long-range order.

\begin{figure}
    \centering
    \includegraphics[width=0.46\textwidth]{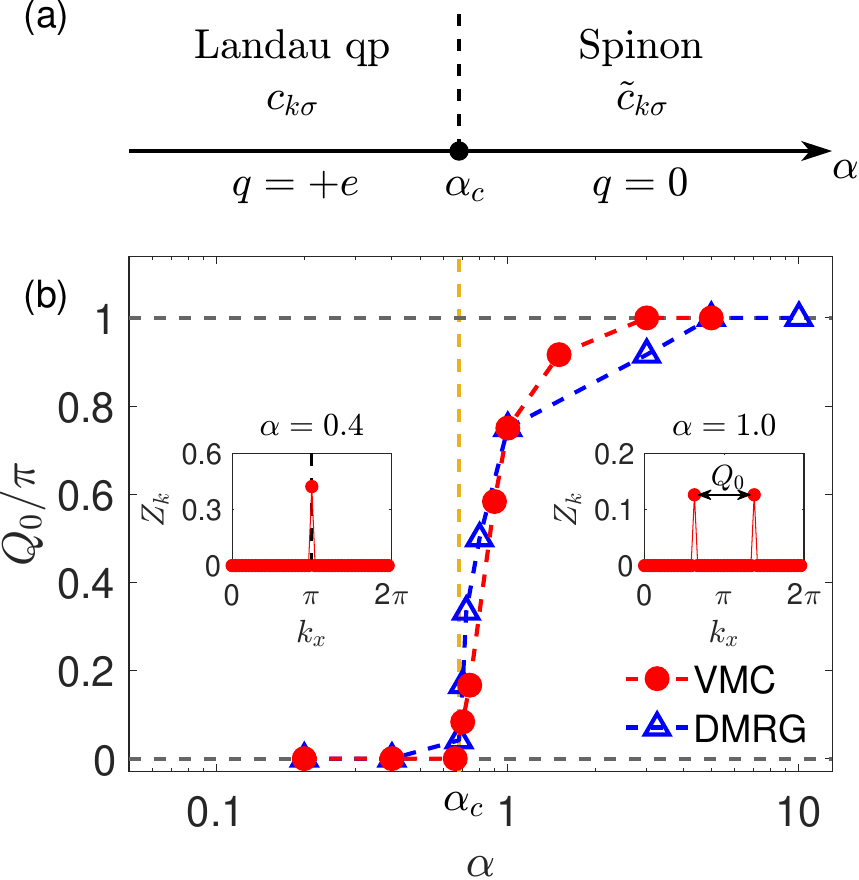}
    \caption{The phase diagram of the single-hole variational ground state as a function of the underlying spin-spin correlation controlled by an anisotropic parameter $\alpha$ in a two-leg $t$-$J$ ladder (see text). (a) A schematic illustration of a quantum transition point at $\alpha=\alpha_c$ for a Landau quasiparticle of charge $q=+e$ state to become a ``twisted'' quasiparticle with diminished charge $q=0$ to be shown in this paper; (b) $Q_0$ [cf. Eq.~\eqref{eqn:CDWQ}] as a function of $\alpha$, which measures the momentum splitting in the ground state (the insets) as calculated by both DMRG and VMC methods on a $48\times 2$ ladder under PBC. The insets show the quasiparticle weight $Z_k$ at two typical $\alpha = 0.4<\alpha_c$ and $\alpha = 1.0 > \alpha_c$, respectively, determined by VMC with $\alpha_c\approx 0.68$ at $t/J=3$. }
    \label{fig:kalpha}
\end{figure}

\begin{figure}
    \centering
    \includegraphics[width=0.3\textwidth]{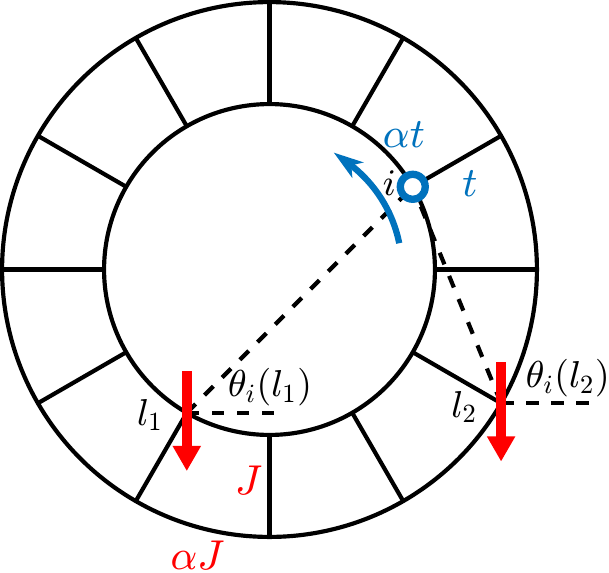}
    \caption{Illustration of a two-leg ladder with anisotropic coupling parameters $t_{ij}$ and $J_{ij}$ of the $t$-$J$ model under the PBC (see text). Here the total number of the ladder sites is $N=N_x\times 2$ with $N_x$ denoting the total number along each of the two legs, which are embedded in 2D with a spatial ring configuration. Note that the two legs of the ladder as rings are with different radii: $r_{\mathrm{in}} = 2-\lambda$ and $r_{\mathrm{out}} = 2+\lambda$, respectively, in which $\lambda$ is a variational parameter to specify the phase-string operator $\hat{\Omega}_i$ of Eq. (\ref{eqn:phasestringoperator}). Here the angle field $\theta_i(l)$, satisfying Eq.~\eqref{theta}, is defined accordingly in the 2D configuration.   }
    \label{fig:PBCphase}
\end{figure}

In this paper, we shall present a simple variational wavefunction ansatz by incorporating the aforementioned phase-string or transverse spin-current effect. It can accurately capture all the essential ground-state features of a single-hole-doped two-leg $t$-$J$ ladder under a periodic boundary condition (PBC) via the VMC method. In particular, it will describe a doped hole as a Landau quasiparticle with charge $q=+e$ in the strong anisotropic limit of $\alpha\ll1$ ($\alpha$ is the ratio of chain direction coupling and rung direction coupling, cf. Fig.~\ref{fig:PBCphase}). In the isotropic regime of $\alpha\simeq 1$, the doped hole 
will acquire nonzero total momenta with finite spin currents, but its charge current will vanish in the large sample limit. The results clearly indicate that the doped hole is a charge-neutral ``spinon'' in the regime of $\alpha>\alpha_c$ with $\alpha_c$ denoting a quantum critical point (QCP), as schematically illustrated in Fig.~\ref{fig:kalpha}(a), which are in excellent agreement with the DRMG calculation [cf. Fig.~\ref{fig:kalpha}(b)].  In other words, we shall demonstrate by an analytic wavefunction that an exotic transition resembling a metal-insulator transition beyond a Landau paradigm \cite{Zhou2013} can happen in a doped Mott insulator. 
The specific form of such a variational wavefunction and its VMC results are briefly outlined as follows. 
 
\subsection{Basic variational results}

Such a single-hole wavefunction ansatz is given by
\begin{equation}
    |\Psi_{\mathrm G}\rangle_{\mathrm{1h}} \propto \sum_{i} e^{i\left(k_0x_i - \hat{\Omega}_i\right) }c_{i\downarrow}|\phi_0\rangle,
    \label{eqn:1hgs}
\end{equation}
where $|\phi_0\rangle$ denotes an undoped spin-singlet background of the two-leg ladder, and the electron annihilation operator $c_{i\downarrow}$ removes an electron of spin $\downarrow$ (without loss of generality) to create a \emph{bare} hole at site $i$. 
The ground state momentum $k_0$ is along the quasi-1D ladder direction, with each rung labeled by $x_i$. 
Note that if one turns off the phase-shift operator $\hat{\Omega}_i$ in Eq.~\eqref{eqn:1hgs}, the wavefunction is simply reduced to a conventional Bloch-wave state as the leading term of a spin polaron wavefunction \cite{Shraiman1988a,Reiter1994},
\begin{equation}
    |\Psi_{\mathrm {B}}(k_0)\rangle_{\mathrm{1h}} \propto \sum_{i} e^{ i k_0 x_i }c_{i\downarrow}|\phi_0\rangle,
    \label{eqn:1hBloch}
\end{equation}
which is explicitly translational invariant (with the translationally invariant spin background $|\phi_0\rangle$) to describe a Landau-like quasiparticle of momentum $k_0$. The phase-shift field $\hat{\Omega}_i$ as a nonlocal spin operator is thus the only unconventional quantity in the ground state of Eq.~(\ref{eqn:1hgs}), which is to incorporate the phase-string effect as stated above. An excellent agreement of the VMC calculation based on Eq.~\eqref{eqn:1hgs} and its variant (i.e., a further incorporation of the longitudinal spin-polaron effect) with the DMRG and ED results will be demonstrated in this paper. 

As a matter of fact, a QCP is found in the ground state as a function of the anisotropic parameter $\alpha$, by which the AF correlation in $|\phi_0\rangle$ can be continuously tuned. 
At strong rung limit ($\alpha \ll 1$), $k_0$ is found at $\pi \mod 2\pi$  (taking the lattice constant as the unit), which corresponds to a non-degenerate state essentially the same as the Bloch-wave state in Eq.~\eqref{eqn:1hBloch}. 
A double-degeneracy in the ground state arises on the other side of the QCP at larger $\alpha$'s, which is characterized by nontrivial momentum splitting $k_0^\pm=\pi \pm \kappa \mod 2\pi$. 
The split as characterized by the wavevector:
\begin{equation}\label{eqn:CDWQ}
    Q_0\equiv 2\kappa 
\end{equation}
is shown in Fig.~\ref{fig:kalpha}(b)
as a function of $\alpha$ calculated by VMC and DMRG, respectively, which both indicate a QCP at $\alpha_c \simeq 0.68$ where $Q_0$ vanishes as determined by DMRG \cite{Zhu2015b,Zhu2015,Zhu2018a,White2015}. 

In the insets of Fig.~\ref{fig:kalpha}(b), the sharp peak(s) of the quasiparticle spectral weight $Z_k$ specifies $k_0$ in the ground state. Here $Z_k$ is defined as the absolute-value squared of the overlap between the wavefunctions in Eqs.~\eqref{eqn:1hgs} and \eqref{eqn:1hBloch} (after normalization) as follows 
\begin{equation} \label{eqn:Zkdefin}
    Z_{k}\equiv \frac{1}{2}\left | \langle\Psi_{\mathrm {B}}(k)|\Psi_{\mathrm G}\rangle_{\mathrm{1h}} \right |^2
    = \left| \langle\phi_0|c^\dagger_{k\downarrow} |\Psi_{\mathrm G}\rangle_{\mathrm{1h}} \right |^2 ~, 
\end{equation}
where $c_{k\downarrow}^\dagger = 1/N\sum_i c_{i\downarrow}^\dagger e^{ikx_i}$ is the $k$ space electron with $k_y=0$. On both sides of the QCP in Fig.~\ref{fig:kalpha}(b), $Z_k$ is always finite, indeed consistent with the DMRG result first shown in Ref.~\onlinecite{White2015}. In particular, the non-degenerate ground state at $\alpha<\alpha_c$ is a Landau quasiparticle, which can be smoothly connected to the Bloch-wave state in Eq.~\eqref{eqn:1hBloch}. 
However, we shall show that the charge of the doped hole will actually \emph{disappear} at $\alpha>\alpha_c$, whereas its spin-1/2 remains unrenormalized. 
In other words, the QCP represents a fundamental transition of the doped hole from a Landau-like quasiparticle to a pure charge-neutral spinon, which is schematically illustrated in Fig.~\ref{fig:kalpha}(a). 
Such a non-Landau-like quasiparticle with a finite $Z_{k_0^\pm}$ indicates a two-component structure in the wavefunction where the Landau's one-to-one correspondence hypothesis fails at $\alpha>\alpha_c$. Indeed, besides a finite amplitude of the Bloch-wave component (with $Z_{k^\pm_0}\neq 0$), another many-body component is also explicitly identified in the ground state, in which a spin current pattern associated with the doped charge is always present. The latter is found to be charge \emph{incoherent} as the total momentum $k_0^\pm$ is now continuously shared between the hole and spin degrees of freedom.

Finally, it is briefly discussed that the pairing between two doped holes also becomes substantially enhanced at $\alpha>\alpha_c$ as previously revealed by the DMRG calculation \cite{Zhu2015b,Zhu2014}. An explicit pairing-mediated spin current pattern is shown based on the present wavefunction ansatz, which illustrates how a strong binding can be indeed realized by eliminating the phase-string effect through the pairing of two holes.

The rest of the paper is organized as follows. 
In Sec.~\ref{sec:model}, we introduce the two-leg anisotropic $t$-$J$ model and construct a single-hole-doped wavefunction ansatz under the PBC. 
A systematical comparison between the DMRG and VMC methods are shown on both sides of the QCP at $\alpha_c$. 
In Sec.~\ref{sec:twist}, the properties of the wavefunction at $\alpha>\alpha_c$  
are further analyzed to show that, different from a Landau quasiparticle, here the ``twisted'' quasiparticle carries a finite spin current in the degenerate ground state but vanishing charge current in the thermodynamic limit.
In Sec.~\ref{sec:discussion}, a further discussion of the underlying physics of the non-Landau quasiparticle behavior is made. In particular, how the incoherent charge component is crucial to the pairing between doped holes is pointed out. Finally, the conclusion and perspectives are given in Sec.~\ref{sec:conclusion}.

\section{Benchmarking Wavefunction Ansatz with ED and DMRG via VMC calculation}\label{sec:model}

\subsection{The two-leg anisotropic $t$-$J$ model}

In this paper, we shall study the single-hole-doped ground state of the $t$-$J$ model on an anisotropic two-leg ladder 
with system size $N = N_x \times 2$. Here the $t$-$J$ Hamiltonian is given by $H = \mathcal{P}_s(H_t+H_J)\mathcal{P}_s$, where 
\begin{align} 
    H_t & =-\sum_{\langle ij\rangle,\sigma}t_{ij}(c_{i\sigma}^\dagger c^{}_{j\sigma}+\mathrm{H.c.})~, 
    \label{eqn:hoppingH} \\ 
    H_J & =\sum_{\langle ij\rangle}J_{ij}\left(\mathbf{S}_i\cdot\mathbf{S}_j-\frac{1}{4}n_in_j\right)~, 
    \label{eqn:superexchangeH}
\end{align}
with $\langle ij\rangle$ denoting a nearest-neighbor (NN) bond. 
Here $\mathbf{S}_i$ and $n_i$ are spin and electron number operators on site $i$, respectively. 
The strong correlation nature of the $t$-$J$ model originates from the 
no double occupancy constraint $\sum_{\sigma} c^\dagger_{i\sigma}c^{}_{i\sigma}\leq 1$ 
on each site, which is imposed via the projection operator $\mathcal{P}_s$. 
Generally, a two-leg ladder is anisotropic along the chain direction (denoted as ${x}$ direction) 
and the rung direction (denoted as $y$ direction) as illustrated in Fig.~\ref{fig:PBCphase} under PBC, where we choose the rung-direction couplings as $t_{ij} = t$ and $J_{ij} = J$, and the chain-direction couplings as $t_{ij} = \alpha t$ and $J_{ij}=\alpha J$, 
with $\alpha>0$ as the anisotropic parameter. The superexchange coupling constant $J$ is taken as the unit and 
the hopping term $t/J=3$ is used throughout the paper. 
The DMRG calculation of this paper is done with 2500 saved states to fit a truncation error up to $10^{-10}$ with 200 sweeps for convergence. 
Most of the VMC calculations in this paper are done on a $48\times 2$ lattice, but no obvious change of the results is seen as the system size changes up to $64\times 2$. 

\subsection{Single-hole wavefunction ansatz}

At half-filling, where the $t$-$J$ model is reduced to the Heisenberg spin model on a bipartite square lattice, 
the ground state is a spin singlet state, 
with a finite spin-gap opened up 
for the two-leg ladder case \cite{Zhu2013}. 
In the following, we shall denote it as $|\phi_0\rangle$.

Then, based on a bare hole state created at site $i$ by removing an electron of spin $\downarrow$ from the spin-singlet background, 
i.e., $c_{i\downarrow}|\phi_0\rangle$, a Bloch-wave-like single-hole state may be constructed as
\begin{equation} \label{eqn:Bloch}
    |\Psi_{\mathrm{B}}\rangle_{\mathrm{1h}} = \sum_{i} \varphi^{\mathrm B}_h (i)c_{i\downarrow}|\phi_0\rangle~,
\end{equation}
where the variational wavefunction $\varphi^{\mathrm B}_h (i)\propto e^{i kx_i}$ is a Bloch-wave with a momentum $k$ along the quasi-1D ladder direction under the translation symmetry. 
In general, the doped hole will induce a many-body response from the spin background, known as the phase-string effect \cite{Sheng1996,Wu2008}, such that the single-hole state can be significantly renormalized beyond the Bloch-wave-like one in Eq.~\eqref{eqn:Bloch}. How to treat such an effect is therefore the central issue in the study of the doped Mott physics.

An ansatz ground state has been previously proposed for the $t$-$J$ model, which is generally given in the one-hole case as follows \cite{Weng2011a,Wang2015,Chen2019} 
\begin{equation}
    |\Psi_{\mathrm{G}}\rangle_{\mathrm{1h}} = \sum_{i} \varphi_h^{}(i)e^{- i\hat{\Omega}_i}c_{i\downarrow}|\phi_0\rangle~,
    \label{eqn:singleholeansatz}
\end{equation}
where a new phase factor $e^{- i\hat{\Omega}_i}$ is explicitly introduced to represent the many-body phase shift or the phase-string effect from the spin background when a hole is created at site $i$. 
In other words, the corresponding spin background is modified from $|\phi_0\rangle$ to $e^{- i\hat{\Omega}_i}|\phi_0\rangle $. 
Here $\varphi_h(i)$ is a variational wavefunction to be optimized, and $\hat{\Omega}_i$ is explicitly given by \cite{Weng2011a,Wang2015,Chen2019}
\begin{equation}
    \hat{\Omega}_i = \sum_{l(\neq i)}\theta_i(l)n_{l\downarrow}~,
    \label{eqn:phasestringoperator}
\end{equation}
where $n_{l\downarrow}$ is the number operator of the down spin at site $l$. 
The statistical angle $\theta_i(l)$ must satisfy the condition
\begin{equation} 
    \theta_i(l)-\theta_l(i)=\pm \pi ~,
    \label{theta}
\end{equation}
for two NN sites $i$ and $l$ such that a sign change can be instantly produced by the phase factor $e^{- i\hat{\Omega}_i}$ when the hole exchanges with a spin of $\sigma =- 1$ during an NN hopping.
As the result, the singular part of the phase-string effect of the $t$-$J$ model can be precisely compensated via $e^{- i\hat{\Omega}_i}$ due to Eq. (\ref{theta}) \cite{Weng2011a,Wang2015,Chen2019}. It is important to point out that the phase-string cannot be truly ``gauged away'' by $e^{- i\hat{\Omega}_i}$, which only serves as a unitary/duality transformation to turn the singular phase-string into a smooth nonlocal/topological effect such that $\varphi_h(i)$ may be still treated as a conventional (Bloch-type) wavefunction (see below).   

Note that Eq. (\ref{theta}) alone does not completely specify $\theta_i(l)$. A simple choice of $\theta_i(l)$ satisfying Eq.~\eqref{theta} may be given by $\theta_i(l)=\pm \mathrm{Im}\ln(z_i-z_l)$ in an isotropic 2D plane \cite{Chen2019}, with $z_i=x_i+iy_i$ being the complex coordinate of site $i$. 
For the two-leg ladder case, previously an anisotropic definition of $\theta_i(l)$ was introduced \cite{Wang2015} with an extra variational parameter. But it can only apply to a finite ladder with an open boundary condition (OBC). 
In the present paper, in order to study the one-hole ground state properties under the PBC, a distinct choice of $\hat{\Omega}_i$ will be needed. Here we shall still use the same isotropic definition of the statistical phase $\theta_i(l)$ given above for the isotropic 2D \cite{Weng2011a,Chen2019} but put the two-leg ladder in a 2D plane with a spatial configuration shown in Fig.~\ref{fig:PBCphase}. With the two legs of the ladder being bent into two rings, the PBC is realized. By making two rings with distinct radii of $r_{\mathrm{in}} = 2 - \lambda$ and $r_{\mathrm{out}} = 2 + \lambda$, respectively, $\lambda$ can be taken as a variational parameter to tune the anisotropy (due to the two-leg instead of 2D) in the phase-string operator $\hat{\Omega}_i$ in place of an anisotropic $\theta_i(l)$ originally defined in Ref.~\onlinecite{Wang2015}. 

It is emphasized that such a detailed spatial configuration in Fig.~\ref{fig:PBCphase} is useful for implementing the phase-string operator $\hat{\Omega}_i$ without changing the model Hamiltonian, which is apparently translational invariant along the ladder direction due to PBC. We point out that some basic important features of the single-hole-doped two-leg ladder have been already well captured by the wavefunction approach in Ref.~\onlinecite{Wang2015}, including the existence of the QCP and the charge modulation pattern in the non-Landau quasiparticle regime. But in contrast to OBC used in Ref.~\onlinecite{Wang2015}, the two-leg ladder under PBC will enable us to study the whole problem under the translational symmetry, where the spin and charge currents, total momentum, and the response under inserting magnetic flux, etc., can be well defined to characterize the ground state. In other words, in the sense that thermodynamic limit is taken and the middle uniform regime of the OBC system is considered, the present PBC approach will consistently cover all the main results obtained under the OBC in Ref.~\onlinecite{Wang2015} and at the same time provide a more complete understanding of the single-hole problem as to be presented below.  

Besides $\lambda$ in the definition of $\hat{\Omega}_i$ [via $\theta_i(l)$ defined in the geometry of Fig.~\ref{fig:PBCphase}], another variational parameter is the single-hole wavefunction $\varphi_h^{}(i)$ in Eq.~(\ref{eqn:singleholeansatz}), which may be determined by optimizing the total ground energy variationally. 
A further simplification is that $\varphi_h^{}(i)$, up to a $U(1)$ phase, is also Bloch-wave-like: $\varphi_h^{}(i)\propto e^{i k_0x_i}$, since generally the spin-spin correlation is short-ranged in $|\phi_0\rangle $ such that the new ``twisted'' hole created by $\tilde{c}_{i\downarrow}\equiv e^{- i\hat{\Omega}_i}c_{i\downarrow}$ should be of a finite scale which is translationally invariant moving on $|\phi_0\rangle $. 
Namely, evaluating the variational parameter of $\varphi_h^{}(i)$ reduces to determining $k_0$, which will become nontrivial in general. 
The detailed VMC procedure is a generalization of Refs.~\cite{Wang2015,Chen2019}, which is outlined in Appendix~\ref{app:3}. 

Finally, the wavefunction ansatz in Eq.~\eqref{eqn:singleholeansatz} can be further improved by introducing a ``longitudinal spin-polaron'' correction to make $|\phi_0\rangle\rightarrow \hat{\Pi}_i|{\phi}_0\rangle $ around the hole site $i$, such that the spin background becomes $e^{- i\hat{\Omega}_i}\hat{\Pi}_i|{\phi}_0\rangle$. Such a procedure will be straightforward as discussed in Appendix~\ref{app:2}.  But as we shall see below, the main body of the physical properties of the single-hole problem is already excellently captured by the simplest form in Eq.~\eqref{eqn:singleholeansatz}, even though some quantitative results can be slightly improved by the extra variational parameters,  such as the total energy and the position of the QCP point $\alpha_c$ (cf. Appendix~\ref{app:2}). 
Physically such a longitudinal spin-polaron effect can serve as a key improvement for the Bloch-wave state in Eq.~\eqref{eqn:Bloch} without changing its nature as a Landau's quasiparticle. 
Similarly, its correction to the present one-hole state in Eq.~\eqref{eqn:singleholeansatz} will not change its non-Landau features either, with only a quantitative improvement in the variational energy as to be shown in the following.

\begin{table}[tb]
    \centering
    \caption{The energies and quantum number of the single-hole variational ground states in comparison with the ED results
        on a $8\times 2$ ladder with $\alpha = 1.0$ and $\alpha = 0.4$: 
        $E_{\mathrm{G}}$ is the total energy; $E_t$ and $E_J$ are the kinetic and the superexchange energies, respectively; 
        $k_0$ denotes the corresponding momentum along the $x$-direction.
        Ground states $|\Psi_{\mathrm{B}}\rangle_{\mathrm{1h}}$ and $|\Psi_{\mathrm{G}}\rangle_{\mathrm{1h}}$ are given in 
        Eqs.~\eqref{eqn:Bloch} and \eqref{eqn:singleholeansatz}, respectively; $|\tilde{\Psi}_{\mathrm{B}}\rangle_{\mathrm{1h}}$ and $|\tilde{\Psi}_{\mathrm{G}}\rangle_{\mathrm{1h}}$ denote the corresponding variational states further improved by incorporating the longitudinal spin-polaron effect (see the main text and Appendix~\ref{app:2}). }
    \begin{ruledtabular}
        \begin{tabular}{lccccc}
            ~& $\alpha$ & $E_{\mathrm{G}}$ & $E_t$ & $E_J$ & $k_0$\\
            \colrule
            $|\Psi_{\mathrm{B}} \rangle_{\mathrm{1h}}$ & $1.0$ & $-15.84$ & $-2.46$ & $-13.37$ & $0$ \\
            $|\tilde{\Psi}_{\mathrm{B}}\rangle_{\mathrm{1h}}$ & $1.0$ & $-18.03$  & $-6.26$ & $-11.77$ & $0$\\
            $|\Psi_{\mathrm{G}}\rangle_{\mathrm{1h}}$ & $1.0$ & $-18.15$  & $-5.55$ & $-12.60$ & $\pm \pi/2$\\
            $|\tilde{\Psi}_{\mathrm{G}}\rangle_{\mathrm{1h}}$ & $1.0$ & $-19.43$  & $-6.96$ & $-12.47$ & $\pm \pi/2$\\
            ED & $1.0$ & $-19.77$  & $-7.30$ & $-12.47$ & $\pm \pi/2$\\
             \colrule
            $|\Psi_{\mathrm{B}} \rangle_{\mathrm{1h}}$ & $0.4$ & $-11.78$ & $-2.90$ & $-8.88$ & $\pi$ \\
            $|\tilde{\Psi}_{\mathrm{B}}\rangle_{\mathrm{1h}}$ & $0.4$ & $-13.03$  & $-4.44$ & $-8.60$ & $\pi$\\
            $|\Psi_{\mathrm{G}}\rangle_{\mathrm{1h}}$ & $0.4$ & $-12.67$  & $-3.89$ & $-8.78$ & $\pi$\\
            $|\tilde{\Psi}_{\mathrm{G}}\rangle_{\mathrm{1h}}$ & $0.4$ & $-13.10$  & $-4.42$ & $-8.69$ & $\pi$\\
            ED & $0.4$ & $-13.14$  & $-4.42$ & $-8.71$ & $\pi$\\
        \end{tabular}	
    \end{ruledtabular}
    \label{tab:relaxlad}
\end{table}

\subsection{Variational ground-state energy}

Based on the wavefunction ansatz $|\Psi_{\mathrm G}\rangle_{\mathrm{1h}}$ in Eq.~\eqref{eqn:singleholeansatz}, the ground state energy, $E_{\mathrm{G}}$, can be determined variationally by the VMC method outlined above. Table~\ref{tab:relaxlad} presents the VMC results of $E_{\mathrm G}$ and the hopping and superexchange energies, $E_t$ and $E_J$, at two typical values of the anisotropic parameter: $\alpha = 1.0$ and $\alpha = 0.4$, respectively, in comparison with the ED results for a $8\times 2$ ladder. In the same table, the corresponding energies of the Bloch-wave state $|\Psi_{\mathrm B}\rangle_{\mathrm{1h}} $ [Eq.~\eqref{eqn:Bloch}] are also given for comparison. 

Table~\ref{tab:relaxlad} shows that the ansatz state $|\Psi_{\mathrm G}\rangle_{\mathrm{1h}}$
gives a much better ground energy than the Bloch-like state $|\Psi_{\mathrm B}\rangle_{\mathrm{1h}}$ at $\alpha = 1.0$.  Most importantly, the ansatz state $|\Psi_{\mathrm G}\rangle_{\mathrm{1h}}$ captures the correct ground state momenta $k_0^\pm = \pm \pi/2 \mod 2\pi$ at $\alpha = 1.0$, which is totally missed by $|\Psi_{\mathrm B}\rangle_{\mathrm{1h}}$. By contrast, at $\alpha=0.4$, both $|\Psi_{\mathrm G}\rangle_{\mathrm{1h}}$ and $|\Psi_{\mathrm B}\rangle_{\mathrm{1h}}$ have the same momentum $k_0=\pi$ and the ground-state energies are also relatively closer even though $E_t$ in the former is still much improved.

According to the phase diagram of Fig.~\ref{fig:kalpha}(a), which is to be elaborated below, the single-hole ground states at $\alpha = 1.0$ and $\alpha = 0.4$ are on the two sides of the critical $\alpha_c\simeq 0.68$, e.g.,  $\alpha = 0.4<\alpha_c$ is in the Landau quasiparticle regime where $|\Psi_{\mathrm G}\rangle_{\mathrm{1h}}$ and $|\Psi_{\mathrm B}\rangle_{\mathrm{1h}}$ may be adiabatically connected, but $\alpha = 1.0>\alpha_c$ is a distinct regime where $|\Psi_{\mathrm G}\rangle_{\mathrm{1h}}$ cannot be reduced to $|\Psi_{\mathrm B}\rangle_{\mathrm{1h}}$ because of a nontrivial ``transverse spin-polaron effect'' due to the phase-shift factor $e^{- i\hat{\Omega}_i}$. 

Before we explore such distinction between the two phases and the quantum transition at $\alpha_c$ below, we further examine an improvement of the ground-state energy by incorporating the aforementioned ``longitudinal spin-polaron effect'', which will turn $|\Psi_{\mathrm G}\rangle_{\mathrm{1h}}$ and $|\Psi_{\mathrm B}\rangle_{\mathrm{1h}}$ into $|\tilde{\Psi}_{\mathrm{G}}\rangle_{\mathrm{1h}}$ and $|\tilde{\Psi}_{\mathrm{B}}\rangle_{\mathrm{1h}}$, respectively (cf. Appendix.~\ref{app:2} for the details). 
The corresponding variational energies on a $8\times 2$ ladder are also shown in Table~\ref{tab:relaxlad}. At $\alpha=1.0$, the ground-state energy of $|\tilde{\Psi}_{\mathrm G}\rangle_{\mathrm{1h}}$ can be optimized to be within the precision of $1.7\%$ as compared to the ED result, with $k_0^\pm $ unchanged. But for the Landau quasiparticle state $|\tilde{\Psi}_{\mathrm B}\rangle_{\mathrm{1h}}$, the energy is still relatively much worse than that of the ansatz wavefunction, with the wrong momentum of $k_0=0$ unchanged.  However, at $\alpha=0.4$, both $|\tilde{\Psi}_{\mathrm{G}}\rangle_{\mathrm{1h}}$ and $|\tilde{\Psi}_{\mathrm{B}}\rangle_{\mathrm{1h}}$ get optimized to be within $0.3\%$ and $0.8\%$ of the exact ED value, consistent with the fact that the longitudinal spin-polaron effect is solely responsible for the renormalization of the doped hole in the Landau quasiparticle regime.

\begin{figure*}
    \centering
    \includegraphics[width=0.86\textwidth]{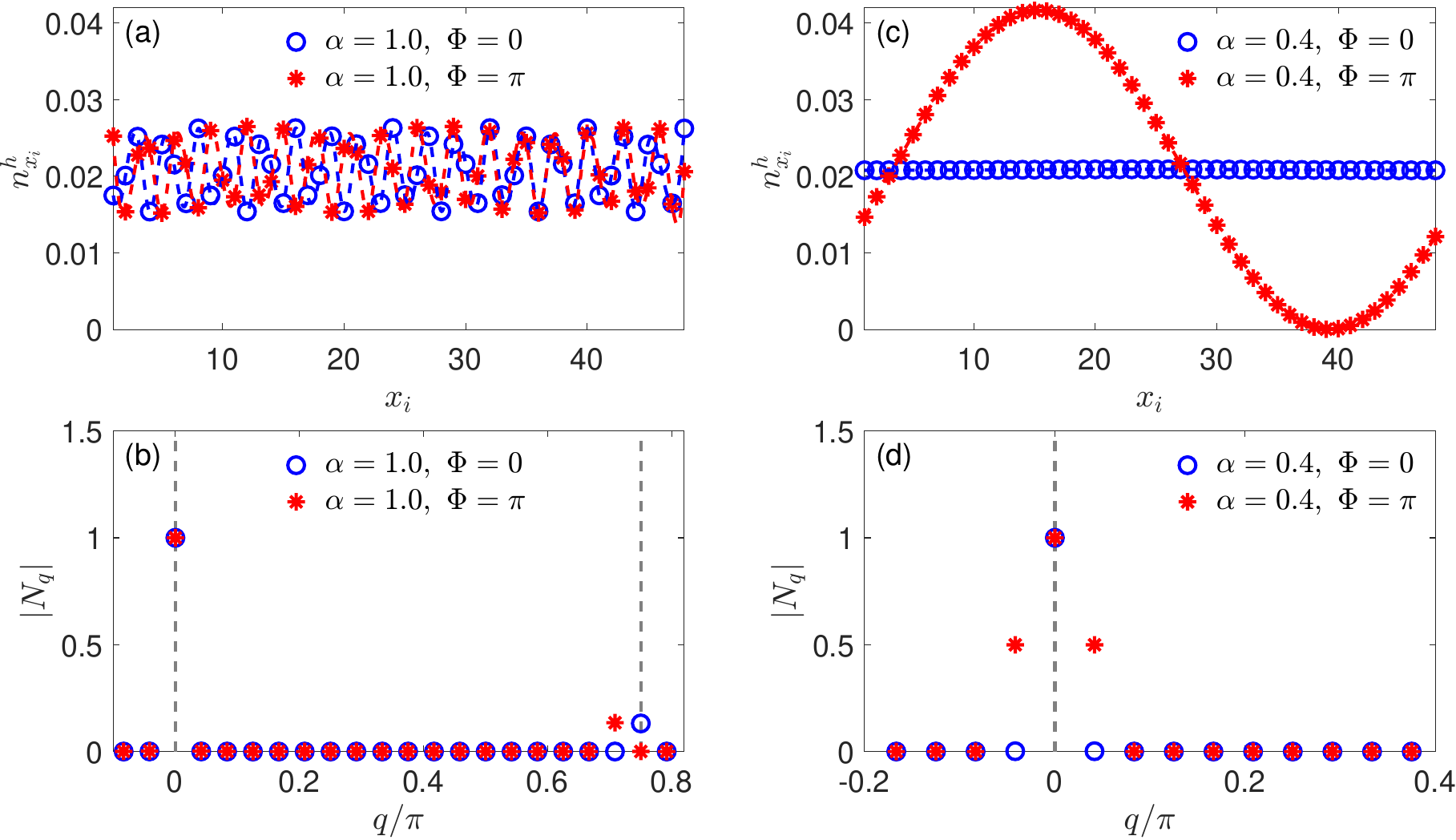}
    \caption{ Ground-state hole density distributions $n_{x_i}^h$ and its Fourier transformation $N_q = \sum_{x_i=1}^{N_x}n_{x_i}^he^{iqx_i}$ 
        calculated by DMRG method on a $48\times 2$ PBC ladder  
        for $\alpha = 1.0 > \alpha_c$ [(a), (b)] and $\alpha = 0.4 < \alpha_c$ [(c), (d)], respectively. 
        The results for a $\Phi=\pi$ flux inserting through the center of the PBC rings are also shown. Note that at $\alpha = 1.0$ 
        the ground states are double degenerate, and a real wavefunction state is used for the measurement. } 
    \label{fig:modulation_dmrg}
\end{figure*}
\begin{figure*}
    \centering
    \includegraphics[width=0.86\textwidth]{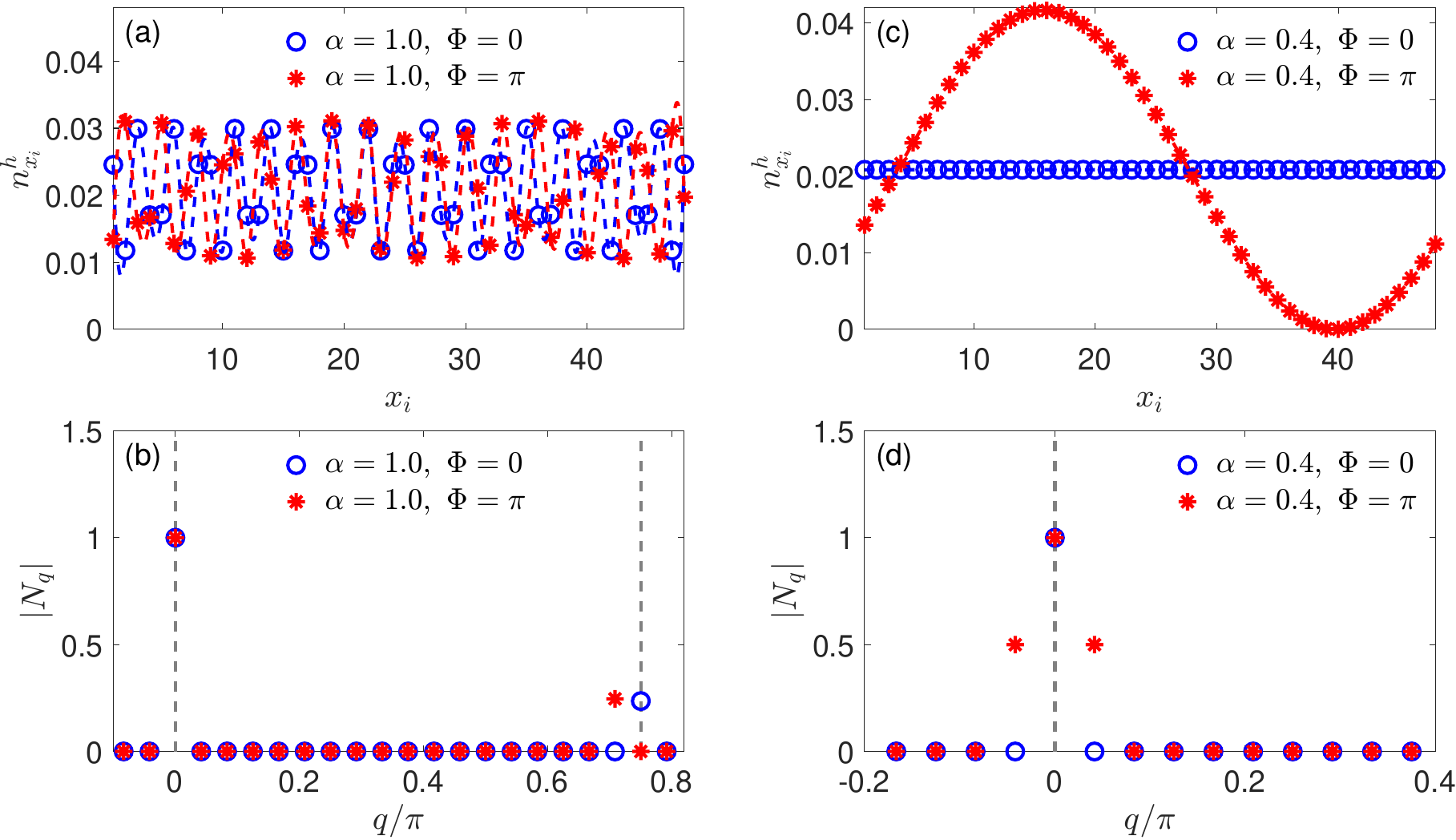}
    \caption{ Ground-state hole density distributions $n_{x_i}^h$ and its Fourier transformation $N_q = \sum_{x_i=1}^{N_x}n_{x_i}^he^{iqx_i}$ 
        calculated by VMC based on the ground-state ansatz in Eq.~\eqref{eqn:singleholeansatz}. The parameters are the same as in Fig.~\ref{fig:modulation_dmrg} for a $48\times 2 $ PBC ladder at $\alpha = 1.0 > \alpha_c$ [(a), (b)] and $\alpha = 0.4 < \alpha_c$ [(c), (d)], respectively. The results for a $\Phi=\pi$ flux inserting through the center of the PBC rings are also shown. }
    \label{fig:modulation_vmc}
\end{figure*}

\subsection{Phase diagram}

The ground-state phase diagram of the single hole doped anisotropic two-leg $t$-$J$ ladder 
has already been carefully studied by DMRG calculations \cite{Zhu2015b,Zhu2015,Zhu2018a,White2015}. 
A key finding is that the analyticity of the ground state energy has a singularity as a function of the anisotropic parameter $\alpha$ 
at $\alpha=\alpha_c\approx 0.68$ ($t/J=3$), which resembles a second-order phase transition \cite{Zhu2015b}. 
Across the critical point $\alpha_c$, the physical properties are also qualitatively changed. For example, the non-degenerate ground state at $\alpha<\alpha_c$ becomes double-degenerate at $\alpha>\alpha_c$ (for a given total spin-1/2). The corresponding hole density distribution changes from flat to a charge modulation characterized by a wavevector $Q_0$ for each of the degenerate (real wavefunction) states, where the wave length $2\pi/Q_0$ is generally incommensurate with the lattice constant $a_0$  at large $N_x$ limit (note that $Q_0=2\pi /a_0\times \mathrm {integer}/N_x$). In Fig.~\ref{fig:kalpha}(b), the critical point $\alpha_c$ is well indicated by the emergence of a finite $Q_0$ as a function of $\alpha$ as calculated by DMRG under PBC for an $N=48\times 2$ ladder.  

\begin{figure}
    \centering
    \includegraphics[width=0.46\textwidth]{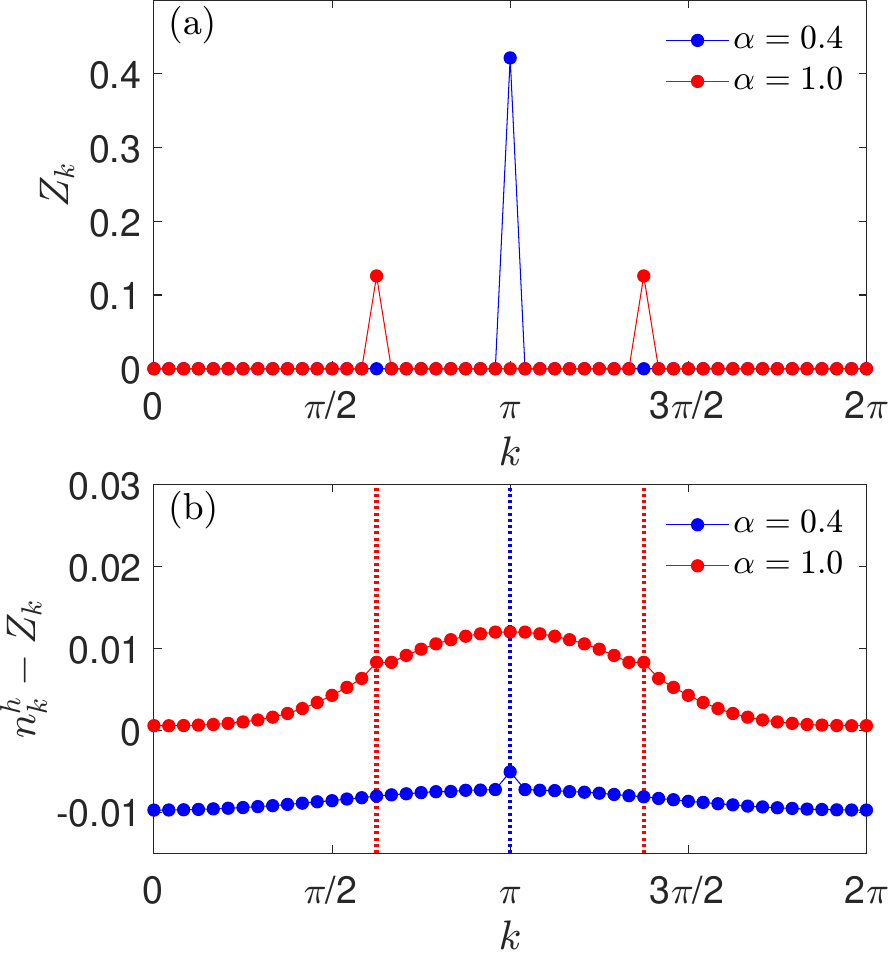}
    \caption{(a) The quasiparticle weight $Z_k$ calculated by VMC at $\alpha = 0.4$ and $\alpha = 1.0$, respectively, on a $48\times 2$ ladder with PBC; 
    (b) Momentum distribution of the hole subtracted by the spectral weight, $n_k^h - Z_k$ . 
    Here the dashed lines mark the positions of the peaks of the quasiparticle weight. For $\alpha>\alpha_c$ a real wavefunction state is used. }
    \label{fig:nhk}
\end{figure}

Such a QCP at $\alpha_c$ can be well quantitatively reproduced based on the ansatz wavefunction given in Eq.~\eqref{eqn:singleholeansatz}.
The VMC result is also presented in Fig.~\ref{fig:kalpha}(b) obtained by optimization with regard to the variational parameters of $\lambda$ and $\varphi_h(i)$, which is then further improved by incorporating the longitudinal spin-polaron effect in $|\phi_0\rangle$ as discussed in the previous subsection (cf.~Appendix~\ref{app:2}). One can see an excellent overall agreement between the VMC and the DMRG result. We mention that a similar QCP has been previously determined under the OBC \cite{Wang2015} based on the singularity in second-derivative of the ground state energy. In the following, to illustrate the quantum transition and distinct behaviors manifested in Fig.~\ref{fig:kalpha}(b), we present two benchmarking calculations, at $\alpha = 1.0 > \alpha_c$ and $\alpha = 0.4 < \alpha_c$, with the DMRG results given in Fig.~\ref{fig:modulation_dmrg} 
and the corresponding VMC results in Fig.~\ref{fig:modulation_vmc}, respectively. 

As comparatively shown in Figs.~\ref{fig:modulation_dmrg}(a) and \ref{fig:modulation_vmc}(a), respectively, the hole average density $n^h_{x_i}$ on a rung $x_i$ of the ladder at $\alpha = 1.0$ ($>\alpha_c$) is presented. 
Here the density profile shows a two-component structure: a flat background plus a modulation with the wavevector $Q_0\neq 0$, with the Fourier transformation along the ladder direction further illustrated in Figs.~\ref{fig:modulation_dmrg}(b) and \ref{fig:modulation_vmc}(b), respectively.
In particular, if a flux $\Phi=\pi$ is inserted into the center of the rings formed by the ladder (cf. Fig.~\ref{fig:PBCphase}), a shift of $Q_0$ by $2\pi/N_x$ is shown for the charge modulation component, while the flat component at the momentum $q=0$ remains unchanged as Figs.~\ref{fig:modulation_dmrg}(b) and \ref{fig:modulation_vmc}(b) indicate.  

By contrast, at $\alpha = 0.4 $ ($< \alpha_c$), a distinct hole profile is given in Figs.~\ref{fig:modulation_dmrg}(c) and \ref{fig:modulation_vmc}(c), whose Fourier transformation is shown in Figs.~\ref{fig:modulation_dmrg}(d) and \ref{fig:modulation_vmc}(d), respectively. They show that the hole profile is uniform with $Q_0=0$. Upon inserting a $\pi$ flux into the rings, however, a node is exhibited in Figs.~\ref{fig:modulation_dmrg}(c) and \ref{fig:modulation_vmc}(c), which corresponds to a momentum shift by $\pm2\pi/N_x$ in Figs.~\ref{fig:modulation_dmrg}(d) and \ref{fig:modulation_vmc}(d). It is consistent with the change of the PBC to an antiperiodic boundary condition (APBC) for a free Bloch wave, indicating that the single-hole state is a Landau quasiparticle which carries a spin-1/2 and charge $+e$ to satisfy both the translation and $U(1)$ symmetries \cite{Zhu2015b,Zhu2018a}.

Such a Landau quasiparticle picture at $\alpha = 0.4$ is consistent with a finite overlap between Eq.~\eqref{eqn:singleholeansatz} and Eq.~\eqref{eqn:Bloch}. As a matter of fact, for $\alpha=0.4<\alpha_c$, a single sharp peak of $Z_{k_0} $ [cf. Eq.~\eqref{eqn:Zkdefin}] at momentum at $k_0=\pi$ is always seen [cf. Fig.~\ref{fig:nhk}(a) or the left inset of Fig.~\ref{fig:kalpha}(b)], which is in agreement with the DMRG \cite{White2015,Zhu2018a}. However, as first indicated \cite{White2015} by DMRG, at $\alpha > \alpha_c$, each of the double-degenerate ground states still has a finite overlap with the Bloch states in Eq.~\eqref{eqn:Bloch}, such that $Z_{k^{\pm}_0}\neq 0$ at momenta $k_0^\pm = \pi \pm \kappa \mod 2\pi$ with $Q_0 = 2\kappa$, as indicated by Fig.~\ref{fig:nhk}(a) or the right inset of Fig.~\ref{fig:kalpha}(b), which is confirmed by the present VMC. Two typical $Z_k$'s at $\alpha=0.4<\alpha_c$ and $\alpha=1.0>\alpha_c$ as determined by VMC are shown in Fig.~\ref{fig:nhk}(a), respectively.  

However, in contrast to the speculation in Ref.~\onlinecite{White2015}, the present ansatz wavefunction will directly show that the single-hole state is no longer a Landau quasiparticle state at $\alpha>\alpha_c$ even though $Z_{k^{\pm}_0}\neq 0$. Based on the ansatz wavefunction in Eq.~\eqref{eqn:singleholeansatz}, one can calculate the momentum distribution of the hole by $n_{k}^h = 1-\sum_\sigma n_{k\sigma}$, where
\begin{equation}
    n_{k\sigma} = \frac{1}{N}\sum_{ij}e^{i{k}(x_i-x_j)}
    \tensor[_{\mathrm{1h}}]{\langle\Psi_G| c^\dagger_{i\sigma}c^{}_{j\sigma}|\Psi_G\rangle}{_{\mathrm{1h}}}~,
    \label{eqn:nhkdef}
\end{equation}
with $k$ the momentum along the chain direction ($k_y=0$). Then, besides $Z_k$ shown in Fig.~\ref{fig:nhk}(a), the VMC results of $n_k^h - Z_k$ are presented in Fig.~\ref{fig:nhk}(b) at $\alpha=0.4$ and $\alpha=1.0$, respectively, which clearly shows a residual broad peak at $\alpha=1.0$, which will be related to an incoherent charge component. In fact, Figs.~\ref{fig:modulation_dmrg}(a) and \ref{fig:modulation_dmrg}(b) and Figs.~\ref{fig:modulation_vmc}(a) and \ref{fig:modulation_vmc}(b) have already indicated the two-component structure of the spatial hole density: the charge modulation with a finite wavevector $Q_0$ and the uniform background. The former responds to inserting a $\pi$ flux into the ring center of the ladder by $2\pi/N_x$ shift in $Q_0$, which is consistent with two quasiparticle components with $Z_{k^{\pm}_0}\neq 0$ at $k_0^\pm = \pi \pm \kappa $,  whereas the latter has no response as it relates to the incoherent background of Fig.~\ref{fig:nhk}(b). In the next section we shall see that such an incoherent part is characterized by a hidden spin current, which continuously carries away momentum from the hole. Note that a rather weak broad peak is also seen $n_k^h - Z_k$ for $\alpha=0.4$ in Fig.~\ref{fig:nhk}(b), which has been argued based on DMRG \cite{Zhu2018a} as completely contributed by the conventional many-body ``cloud'' effect of a Landau quasiparticle. In fact, it does not lead to an incoherent component as shown in Figs.~\ref{fig:modulation_dmrg}(c) and \ref{fig:modulation_dmrg}(d) and Figs.~\ref{fig:modulation_vmc}(c) and \ref{fig:modulation_vmc}(d). Furthermore, the hidden spin current will be completely absent at $\alpha <\alpha_c$ as to be discussed later.

\section{Further characterization:  Novel charge renormalization at $\alpha>\alpha_c$} \label{sec:twist}

In the last section, the wavefunction ansatz based on Eq.~\eqref{eqn:singleholeansatz} has been shown to very accurately capture the essential physics of the single-hole ground state at both $\alpha > \alpha_c$ and $\alpha<\alpha_c$, including the critical point $\alpha_c$, in comparison with both DMRG and ED results. 
In the following, we shall focus on the unconventional case at $\alpha > \alpha_c$ and demonstrate that the charge carried by the doped hole is renormalized to zero. In other words, the QCP at $\alpha_c$ marks the transition of the doped hole from a Landau quasiparticle to a charge-neutral spinon.  

\subsection{Emergent spin-current around the hole at $\alpha > \alpha_c$: Incoherent charge component}
\label{subsec:spincur}

Let us start by emphasizing that the double-degenerate ground states at $\alpha > \alpha_c$ have a two-component structure. 
In the real wavefunction description as shown by DMRG and VMC in Figs.~\ref{fig:modulation_dmrg} and \ref{fig:modulation_vmc}, respectively, the spatial distribution of the doped hole has
a charge modulation component composed of two Bloch-like waves with momenta $k_0^{\pm} = \pi \pm \kappa$ and a uniform background.  The latter is not sensitive to the change of the boundary condition via inserting a flux into the ring of the ladder and has been argued to be an incoherent charge component, which we shall examine more carefully below.  

\begin{figure}
    \centering
    \includegraphics[width=0.46\textwidth]{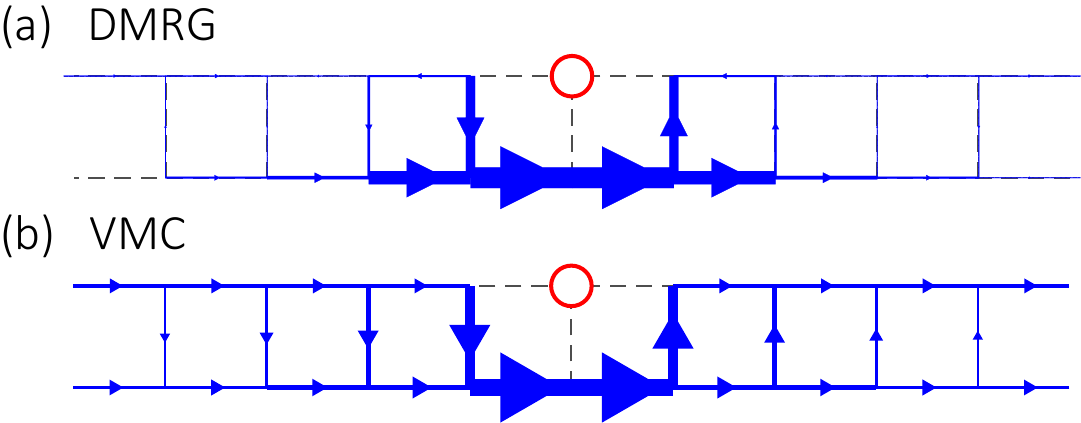}
    \caption{Neutral spin current $J_{ij}^s$ [cf. Eq.~\eqref{eqn:spincurrent1}] (blue arrows) surrounding the hole, which is projected onto a fixed position (read circle) at momentum $k^-_0=\pi-\kappa$ for $\alpha = 1.0$. (a) DMRG on a $24\times 2$ ladder and (b) VMC on a $48\times 2$ ladder. The thickness of the blue lines represents the strength of the spin current, which conserves $S^z$. }
    \label{fig:spincurrent}
\end{figure}

One may also focus on a translational invariant ground state with a total momentum, say, 
$k_0^-=\pi - \kappa$, which has a finite overlap with the Bloch-wave state in Eq.~\eqref{eqn:1hBloch} 
at the given $k=k_0^-$. Then the incoherent charge component in the ground state can be distinguished from the Landau quasiparticle component by the presence of a spin current pattern around the hole, which is 
shown in Figs.~\ref{fig:spincurrent} as calculated by (a) DMRG and (b) VMC at $\alpha =1.0$, respectively.
Here the neutral spin-current operator $J^s_{ij}$, which conserves $S^z$, is defined on the NN link of $i$ and $j$ by 
\begin{equation} 
    J_{ij}^s =i\frac{J}{2}(S^+_iS^-_j-S^-_iS^+_j)~,  \label{eqn:spincurrent1}
\end{equation}
while the backflow spin current of the hopping term is zero as the hole is projected onto a given site (marked by open red circle) as shown in Fig.~\ref{fig:spincurrent}. 
Since the Landau component (Bloch-wave) does not contribute to $J_{ij}^s $, the spin current pattern in Fig.~\ref{fig:spincurrent} 
should entirely come from the non-Landau-component in the ground state. The above results are also consistent with the chiral spin currents observed previously 
for the single-hole problem in the 2D $t$-$J$ systems by ED and DMRG \cite{Zheng2018b} and verified by VMC \cite{Chen2019,Zhao2022}. 

\begin{figure}
    \centering
    \includegraphics[width=0.46\textwidth]{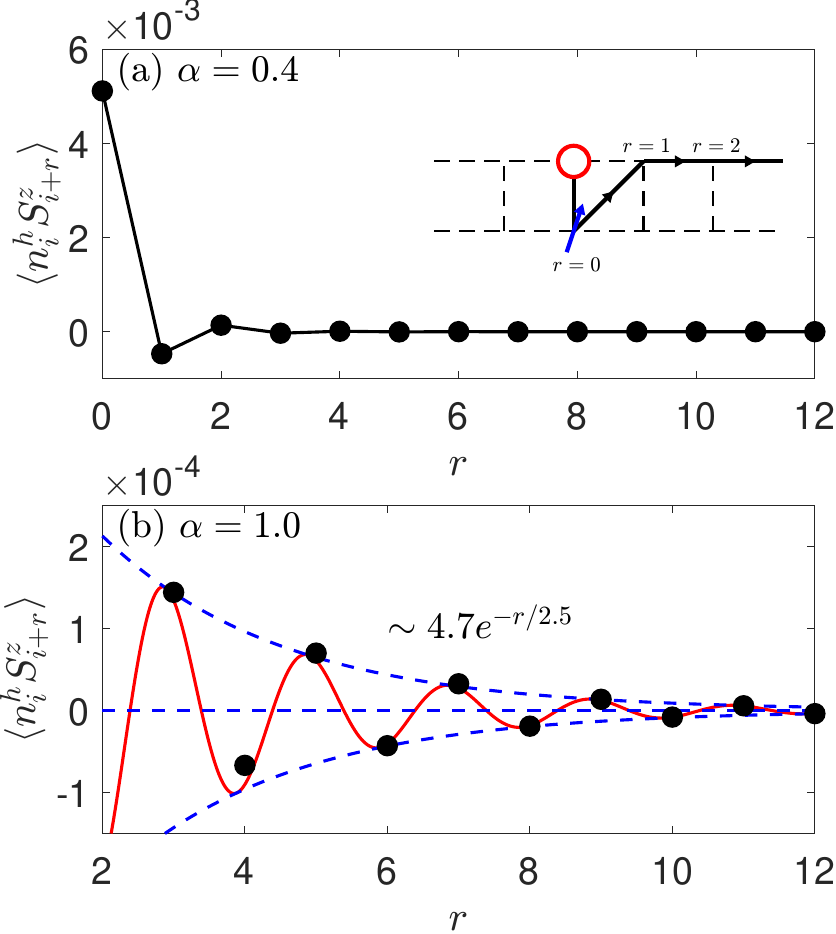}
    \caption{Spin-charge correlator $\langle n_i^hS^z_{i+r}\rangle $ (solid circle) calculated by VMC on a $48\times 2$ ladder. It shows a qualitative change from a tight hole-spin binding to a loosely-bound hole-spin composite from (a) $\alpha = 0.4 < \alpha_c$ to (b) $\alpha = 1.0 > \alpha_c$ in agreement with the DMRG result \cite{Zhu2015b,White2015}. The label $r$ is defined in the inset of (a). The red line and the dashed line in (b) are the fitted curve and envelope function. }
    \label{fig:nhsiz}
\end{figure}

So the spin currents around the hole indicate that the spin partner $S^z=1/2$ is only loosely bound to the hole, which forms a spin current vortex in the hole composite. On the other hand, no spin currents are seen by both VMC and DMRG methods at $\alpha<\alpha_c$, where the non-Landau-component disappears completely. The ``longitudinal spin-polaron'' correction here does not create a transverse spin current surrounding the hole. The absence of the spin currents can be understood as that the spin partner is tightly bound to the doped hole in the strong rung limit to form a conventional Landau quasiparticle with charge $+e$ and spin-1/2.
 
This picture can be also confirmed by measuring the spin-charge correlator $\langle n_i^h S_j^z\rangle$, 
which characterizes the relative distance between the doped hole and an unpaired spin-1/2. 
The results calculated by VMC are shown in Fig.~\ref{fig:nhsiz} for both $\alpha = 0.4$ and $\alpha = 1.0$, respectively.
At $\alpha=0.4$, the spin-1/2 and the hole are tightly bound at the same rung as $\langle n_i^h S_j^z\rangle$ decays quickly at $r>1$ [cf. Fig.~\ref{fig:nhsiz} (a)]. 
On the other hand, the unpaired spin is only loosely bound with the doped hole at $\alpha = 1.0$ as shown in Fig.~\ref{fig:nhsiz} (b). Both behaviors are again in good agreement with the DMRG results \cite{Zhu2015b,White2015}.

\begin{figure}
    \centering
    \includegraphics[width=0.46\textwidth]{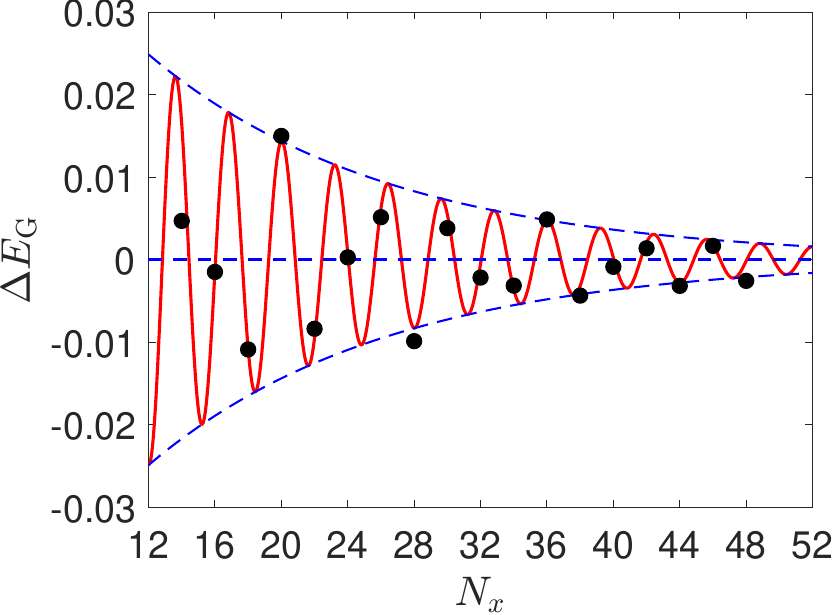}
    \caption{The ground-state energy change $\Delta E_{\mathrm G}$ under a $\pi$ flux insertion ladder as a function of lattice size $N_x$, which is determined by VMC at $\alpha=1.0$. The red-solid line is a fitted curve whose equation is given in Eq.~\eqref{eqn:fitteddElx} and the blue dashed lines are the envelope function of an exponential decay with a correlation length $\xi \approx 14.6$ in agreement with the DMRG result \cite{Zhu2013}.}
    \label{fig:depi}
\end{figure}

\subsection{Response to external electromagnetic flux}
\label{subsec:dedpi}

Even though the single-hole ground state at $\alpha > \alpha_c$ still has a finite overlap with the Bloch-wave state, i.e., $Z_{k_0}\neq 0$, the presence of an intrinsic \emph{incoherent} component makes the Landau's one-to-one hypothesis invalid as the latter cannot be completely specified by a total momentum $k_0$ alone. 
In the above, one has seen that the hopping of the hole will always generate a spin current via the backflow, which means that the momentum $k_0$ is now shared between the (bare) hole and its $S=1/2$ partner, which result in a broad momentum distribution of the bare hole as has been carefully examined by DMRG \cite{Zhu2018a}. In other words, the origin of the incoherent component comes from the internal relative motion inside the loosely-bound hole-spin composite at $\alpha > \alpha_c$.    

The non-Landau behavior may be understood as follows. Because of the finite off-diagonal transition between the coherent and incoherent components, the hole composite as a single entity cannot exhibit a definite charge in response to an external electromagnetic field. Imaging that such a hole circles through the ladder once with a flux $\Phi$ inserting in the hole of the rings formed by the ladder. A charge $q$ will pick up a Berry phase $q\Phi$, which has been well demonstrated by DMRG \cite{Zhu2015b} at $\alpha <\alpha_c$, but an exponentially diminished effect of the flux with the increase of the ladder length $N_x$ has been seen at $\alpha >\alpha_c$ \cite{Zhu2013,Zhu2015b}. In the following, we repeat the calculation of the same effect based on the wavefunction ansatz in Eq.~\eqref{eqn:singleholeansatz}. 

In Sec.~\ref{sec:model}, the response of the charge density distribution to an external $\Phi=\pi$ flux insertion has already been shown. The vanishing charge response for the incoherent component is clearly demonstrated there. Here we calculate the change of the ground-state energy $E_{\mathrm G}$ under the flux insertion: $\Delta E_{\mathrm G} \equiv E_{\mathrm G}(\Phi=\pi) - E_{\mathrm G}(\Phi=0)$. 
Figure~\ref{fig:depi} shows $\Delta E_{\mathrm G}$ as a function of the ladder length $N_x$ calculated by VMC at $\alpha = 1.0$. The black dots are the calculated data, which is fitted by the red-solid curve with the following expression: 
\begin{equation}
    \Delta E_{\mathrm G} = \Delta E_0 e^{-N_x/\xi}\cos(k_0 N_x + \phi_E)~,
    \label{eqn:fitteddElx}
\end{equation}
where $k_0$ is the ground-state momentum at $\alpha = 1.0$, $\Delta E_0$, $\xi$ and, $\phi_E$ are fitting parameters. 
The blue dashed curves are the envelope functions of Eq.~\eqref{eqn:fitteddElx}, which show an exponential decay behaviors $\pm \Delta E_0e^{-N_x/\xi}$ with a length scale $\xi\approx 14.6$. It agrees with a previous calculation by DMRG \cite{Zhu2013} with an exponential decay $\xi\approx 14.5$.

\begin{figure}
    \centering
    \includegraphics[width=0.46\textwidth]{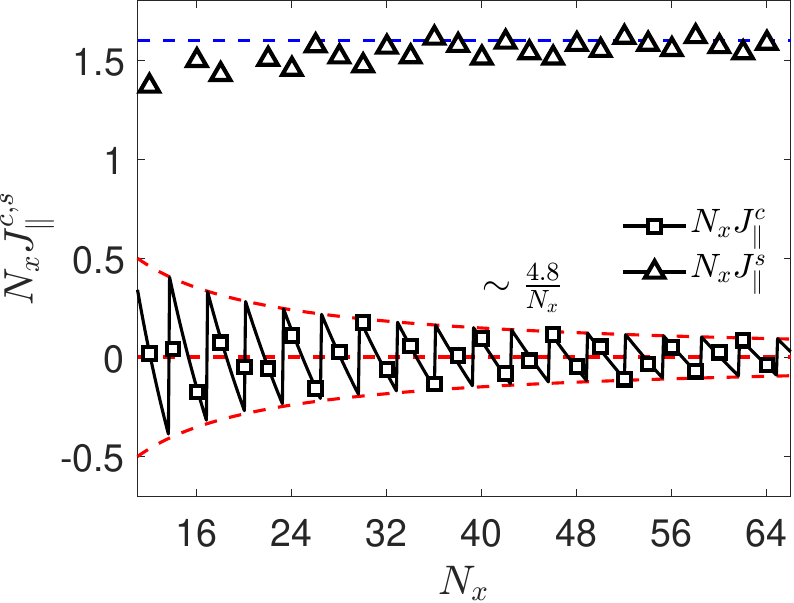}
    \caption{The doped hole becomes a novel spinon at $\alpha=1.0>\alpha_c$: the scaling behaviors of the neutral spin current $N_xJ^{s}_\parallel$ (open triangular) and the charge current $N_xJ^c_\parallel$ (open square) along the ladder direction for a doped hole with given momentum $k^{\pm}_0$. The charge current oscillates and decays to zero, while the neutral spin current saturates to a fixed value (dashed blue) in the long ladder limit. The solid-black line is a fitted curve given in Eq.~\eqref{eqn:chargecurfit} with the red-dashed line as the envelope function.  }
    \label{fig:scalingcur}
\end{figure}

\subsection{Vanishing charge}
\label{subsec:vacharge}

The vanishing response to an external electromagnetic field can be seen in a more explicit way by directly measuring the charge currents of the ground state, 
\begin{equation}\label{eqn:chargecurrent}
    J^c_{ij} = iqt \sum_\sigma (c^\dagger_{i\sigma}c^{}_{j\sigma} - c^\dagger_{j\sigma}c^{}_{i\sigma})~,
\end{equation}
where $i,j$ are NN bonds of the ladder direction and $q = +e$ is the charge of a bare hole. 
For a given momentum $k_0$, a Landau quasiparticle with charge $q$ is expected to carry a finite charge current $J^c_\parallel$ along the ladder direction 
\begin{equation}\label{eqn:finitechargecur}
    J^c_\parallel\equiv 2J_{i,i+\hat{x}}^c=\rho\frac{q}{m^*}\sin k_0 ~,
\end{equation}
where $\rho = 1/N_x$ is the density of a single doped hole on every rung of the ladder, $m^*$ is the effective mass of the quasiparticle, which is found finite on both sides of the QCP \cite{Zhu2015b}, 
and $j=i\pm \hat{x}$ is an NN site of $i$ along the chain direction. 
On the other hand, the neutral spin current of the spin background is expected to vanish 
\begin{equation}\label{eqn:vanishingneutral}
    J^s_\parallel\equiv 2J^s_{i,i+\hat{x}}=0 ~. 
\end{equation}

The above results can be easily verified at $\alpha <\alpha_c$ (for a low-lying excited state with $k_0$ deviates from $\pi$). 
However, at $\alpha =1.0>\alpha_c$, 
one finds that the currents associated with the doped hole of momentum $k_0^-=\pi-\kappa$ are dramatically changed as shown in Fig.~\ref{fig:scalingcur}. 
Instead of the finite charge current $J^c_\parallel/\rho = N_x J^c_\parallel$ as predicted by Eq.~\eqref{eqn:finitechargecur}, 
an oscillating decay behavior of the charge currents is seen, satisfying a fitted curve 
\begin{equation}\label{eqn:chargecurfit}
    N_xJ^c_\parallel = \left. 9.6 \left(\left\{ \frac{k_0}{2\pi}(N_x-\phi_J)+0.5 \right\}-0.5\right)\middle/N_x^{0.94}\right.~,
\end{equation}
where the curly bracket in the expression takes the decimal part of the number inside it, 
$k_0$ is the momentum of the ground state, and $\phi_J$ is a phase parameter to be optimized. 
The overall $N_x^{0.94}$ in the denominator approximate a $1/r$ decay of the envelope function. 

On the other hand, the neutral spin current $N_xJ^s_\parallel$ in the spin background (not the spin current associated with the hopping of the hole), 
which is expected to vanish in the quasiparticle picture as Eq.~\eqref{eqn:vanishingneutral}, 
saturates to a finite value $N_x J^s_\parallel \sim 1.6J$ instead. 
These results confirm that the doped hole is no longer a charged object. 
Instead, the finite neutral spin current $J^s_\parallel/\rho$ indicate that the true quasiparticle is now a spinon. 

\section{Discussion}\label{sec:discussion}

As described in the previous sections, the systematic agreements between the VMC and exact numerical calculations indicate that the wavefunction ansatz of Eq.~\eqref{eqn:1hgs} has reasonably captured all the fundamental physics of the single-hole ground state in the two-leg ladder. It therefore enables one to further examine the underlying mechanism based on the analytic structure of the wavefunction. 

The sole distinction between the wavefunction in Eq.~\eqref{eqn:1hgs} and a conventional bare-hole state in Eq.~(\ref{eqn:1hBloch}) lies in the phase-string factor $e^{-i\hat{\Omega}_i}$, which contributes to a many-body phase shift to the quasiparticle wavefunction $\varphi_h(i) \propto e^{ik_0x_i}$. Here $\hat{\Omega}_i$ involves the background spins nonlocally according to the definition given in Eq.~\eqref{eqn:phasestringoperator}, which is apparently nonperturbative in nature. Since the undoped spin system $|\phi_0\rangle$ is short-range-AF correlated in the two-leg ladder, the exotic quantum entanglement between the doped hole with the surrounding spins are also short-ranged. In particular, the AF correlation length along the ladder direction can be continuously tuned by the anisotropic parameter $\alpha$, which results in a quantum transition at $\alpha_c$ as shown in Fig.~\ref{fig:kalpha}.

\subsection{QCP at $\alpha_c$}

At the QCP, the ground-state momentum $k_0$ splits from $\pi$ to $\pi\pm \kappa$. The incommensurate wavevector $Q_0\equiv 2\kappa$ shown in Fig.~\ref{fig:kalpha} is directly related to the phase-shift operator $\hat{\Omega}_i$ in Eq.~\eqref{eqn:1hgs}. As a matter of fact, based on Eq.~\eqref{eqn:1hBloch} without $\hat{\Omega}_i$, one always finds $k_0=\pi$ without $\alpha_c$.

In the strong-rung limit $\alpha\ll 1$, the singlet-pairing of spins in $|\phi_0\rangle$ is mainly concentrated along rungs. 
In this limit $e^{- i\hat{\Omega}_i}$ is ineffective along the ladder such that the ansatz state \eqref{eqn:1hgs} reduces to the Bloch-wave one in Eq.~\eqref{eqn:1hBloch} at a single momentum $k_0=\pi$.
It has been shown that a longitudinal spin-polaron correction may further improve the ground-state energy, but the hole as a rigid entity of spin-1/2 and charge $+e$ with the same $k_0$ remains robust, which thus satisfies the one-to-one correspondence principle for a Landau quasiparticle, so long as the spin-spin correlation is sufficiently short (than a lattice constant) along the quasi-1D direction.

In the opposite limit of $\alpha\gg 1$, the longer-range spin singlet pairing along the chain direction become more and more important. 
As previously shown, in the 1D case, $e^{- i\hat{\Omega}_i}$ will play a crucial role to result in a momentum at $k_0=\pm \pi/2$  \cite{Zhu2016}. Indeed $Q_0\rightarrow \pi$ at $\alpha\gg \alpha_c$ as shown in Fig.~\ref{fig:kalpha}(a).  

Thus,  $\alpha=\alpha_c$ is the point where $e^{- i\hat{\Omega}_i}$ starts to play a nontrivial role. Note that because of the spin-singlet pairing, $|\phi_0\rangle$ itself does not significantly contribute to a phase shift in $e^{- i\hat{\Omega}_i}$. But in the bare hole state $c_{i\downarrow}|\phi_0\rangle$ with removing a spin $\downarrow$ at site $i$, a spin from the original singlet pair in $|\phi_0\rangle$ will be left unpaired, which in general can make a nontrivial contribution to the momentum shift via $e^{- i\hat{\Omega}_i}$ acting on $c_{i\downarrow}|\phi_0\rangle$. 
In the limit of $\alpha\ll 1$, this unscreened spin mainly stays at the rung direction across the hole as already shown in Fig.~\ref{fig:nhsiz}(a), 
which does not contribute to a momentum shift along the ladder direction. 
At $\alpha>\alpha_c$, the unpaired spin (spinon) is loosely separated from the hole along the ladder direction, as previously shown in Fig.~\ref{fig:nhsiz}(b). 
Due to the presence of $e^{- i\hat{\Omega}_i}$, such a composite effect is clearly manifested by a chiral spin current surrounding the hole in Fig.~\ref{fig:spincurrent}, which contributes to the finite momentum splitting, i.e., $k^\pm_0=\pi\pm \kappa$ with $Q_0\neq 0$. The VMC calculation shows a self-consistent procedure, which minimizes the total energy,  giving rise to a QCP at $ \alpha_c\neq 0$ and the total momentum $k_0$ as a function of $\alpha$. One may find some more detailed account of the underlying mechanism in Appendix.~\ref{app:1}.

\subsection{The two-component structure at $\alpha>\alpha_c$}

We have seen that the double-degenerate ground state at $\alpha>\alpha_c$ must be characterized by a two-component structure, i.e., a Landau-like quasiparticle with $Z_{k_0^\pm }\neq 0$, charge $+e$, and spin $1/2$, and an incoherent component with $Z_{k_0^\pm}=0$ and charge $0$. The latter exhibits  a chiral spin current pattern around the hole, which cannot be produced by the quasiparticle component as shown in Sec.~\ref{subsec:spincur}.  Generally for the left-moving and right-moving hole states at momenta $k_0^\pm$, the chiralities of the spin currents are opposite. In a real wavefunction representation discussed in Sec.~\ref{sec:model} D by a superposition of the $k_0^+ = \pi + \kappa$ and $k_0^- = \pi - \kappa$ states, the incoherent component corresponds to a charge uniform component with no response to an external inserting flux. 

Let us first consider a conventional Landau quasiparticle with momenta $k_0^\pm = \pi \pm \kappa$. 
Mathematically, a real-wavefunction state as a superposition of the $\pi + \kappa$ and $\pi - \kappa$ states may take the form 
\begin{equation} \label{eqn:landaukmk}
    |\psi_{\mathrm{B}}\rangle_{\mathrm{1h}} \propto \sum_i(-1)^{x_i}\left(e^{-i{\kappa} x_i} + \mathrm{c.c}.  \right)c_{i\downarrow}\hat{\Pi}_i|\phi_0\rangle~, 
\end{equation}
where $\hat{\Pi}_i$ is a local spin-polaron operator around the doped hole, which is translationally invariant. Equation ~\eqref{eqn:landaukmk} will result in a hole density modulation by
\begin{equation}
    n_{x_i}^h \propto \sum_{y_i} \langle \phi_0|\hat{\Pi}^{\dagger}_in_{i\downarrow}\hat{\Pi}_i|\phi_0\rangle(1+\cos(2\kappa x_i))~,
    \label{eqn:landaunk}
\end{equation}
which has nodes at $1+\cos(2\kappa x_i) = 0$ with $\kappa \neq 0$. 

In contrast, for the wavefunction ansatz in Eq.~\eqref{eqn:1hgs} with a nontrivial phase factor $e^{-i\hat{\Omega}_i}$. By similarly constructing a real-wavefunction state via a superposition of $k_0^\pm = \pi \pm \kappa$ states as follows 
\begin{equation} \label{eqn:twistkmk}
    |\psi_{\mathrm{G}}\rangle_{\mathrm{1h}}\propto \sum_i(-1)^{x_i}\left(e^{-i{\kappa}{x}_i - i\hat{\Omega}_i}+ \mathrm{c.c}.  \right)c_{i\downarrow}|\phi_0\rangle~,
\end{equation}
one finds a hole density modulation given by 
\begin{equation}
    n_{x_i}^h \propto \sum_{y_i}(\langle \phi_0|n_{i\downarrow}|\phi_0\rangle + 
    \langle \phi_0|n_{i\downarrow}\cos(2\kappa x_i + 2\hat{\Omega}_i)|\phi_0\rangle)~.
    \label{eqn:twistnk}
\end{equation}
The first term is just the uniform density background shown Figs.~\ref{fig:modulation_dmrg}(a) and \ref{fig:modulation_vmc}(a), while the second term corresponds to the modulation part. (For simplicity the longitudinal spin-polaron effect 
has been set as $\hat{\Pi}_i=1$ here.) Different from the modulation of a quasiparticle in Eq.~\eqref{eqn:landaunk}, 
the phase factor of modulation part in the cosine function is scrambled by the phase operator $\hat{\Omega}_i$. 
As a result, it is smaller than the first term because of the inequality 
\begin{equation}
    |\langle \phi_0|n_{i\sigma}e^{-2i\hat{\Omega}_i}|\phi_0\rangle| 
    \leq |\langle \phi_0|n_{i\sigma}|\phi_0\rangle|~, 
    \label{eqn:overlapchiral}
\end{equation}
which leads to a nodeless charge modulation at $\alpha>\alpha_c$ (cf. Fig. \ref{fig:modulation_vmc} in Sec.~\ref{sec:model} D).

\subsection{Disappearance of charge response to external magnetic flux at $\alpha>\alpha_c$}

A simple Landau-like quasiparticle state at small $\alpha$ experiences a quantum phase transition to a novel quasiparticle as the AF correlation length is increased, leading to the nontrivial (incommensurate) momenta in the ground state at $\alpha>\alpha_c$. In a sharp contrast to a conventional Landau-like quasiparticle with a finite momentum, the new particle no longer carries a finite charge current like the former, even though it still carries a spin current contributed by spin-1/2 as a quasiparticle, i.e., a spinon, as shown in Sec.~\ref{sec:twist}. 

To understand how the charge current of twisted quasi-particle $\tilde{c}_{i\sigma}$ is renormalized to zero, one may inspect the averaged 
charge current $J^c_\parallel$ in terms of the ground state of Eq.~\eqref{eqn:singleholeansatz}, 
\begin{equation}
    \begin{aligned}
        J^c_\parallel = &\sum_{y_i}\tensor[_{\mathrm{1h}}]{\langle \Psi_{\mathrm{G}}|J^c_{i,j}|\Psi_{\mathrm{G}}\rangle}{_{\mathrm{1h}}} \\
        \sim & \frac{\sigma t}{N_x}\langle \phi_0|
        \sin({k}_0^\pm\mp( \hat{A}_{ij}^s - \phi^0_{ij})) (n_{i\bar{\sigma}}n_{j\bar{\sigma}}-S_i^\sigma S_j^{\bar{\sigma}}) |\phi_0\rangle~,
    \end{aligned}
\end{equation}
where we have used $\varphi_h(i)\sim \frac{1}{\sqrt{2N_x}}e^{ik_0^\pm x_i}$. 
In addition to the phase factor $e^{ik_0^\pm}$ contributed by the variational parameter, 
a gauge field induced by the spins $\hat{A}^s_{ij} = \sum_{l(\neq i,j)}(\theta_{i}(l)-\theta_j(l))S_l^z$, 
and a uniform $\pi$ flux $\phi^0_{ij} = \frac{1}{2} \sum_{l(\neq i,j)}(\theta_{i}(l)-\theta_j(l))$ emerges from the phase factor $e^{-i\hat{\Omega}_i}$. 
For the present single-hole doped case where a single unpaired spin $S = 1/2$ is present in the spin background, the local flux fluctuation of $\hat{A}^s_{ij}$ is large ($\sim \pi$) as the unpaired single spin is not tightly bound to the doped hole along the ladder direction at $\alpha > \alpha_c$. 
As the result, the charge current contributed by the finite momentum wavefunction $\mathrm{Im}(e^{ik_0^\pm})$ 
is well compensated by the \emph{internal} gauge field generated by the background spins, i.e., $\hat{A}^s_{ij}$.

In the same spirit, the strong fluctuating gauge field $\hat{A}^s_{ij}$ (with strength $\sim \pi$) can effectively screen out 
a weak effect from the external electromagnetic field $A^e_{ij}$. 
Indeed, an exponential decaying behavior in the response to inserting an external $\Phi = \pi$ flux has been shown in Fig.~\ref{fig:depi}, 
which is in contrast to the power law ($1/N_x^2$) behavior of a charged Landau quasiparticle state at $\alpha<\alpha_c$.

\begin{figure}
    \centering
    \includegraphics[width=0.46\textwidth]{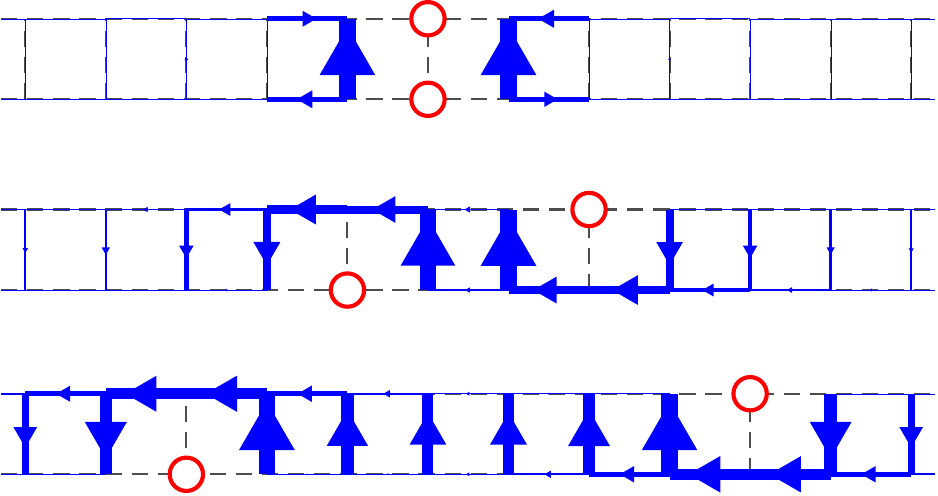}
    \caption{Neutral spin currents with on a $40\times 2$ OBC ladder with $\alpha=1.0$. 
        The two holes are projected at the red circle positions. 
        While the spin currents is indicated by the blue arrows. 
        The strength of the spin currents is represented by the thickness of the arrows. }
    \label{fig:spincurrent2h}
\end{figure}

\subsection{Binding force between two doped holes}

Finally, we briefly discuss the ground state of two doped holes in the two-leg ladder based on the present single-hole wavefunction. Previously the DMRG calculation has also indicated \cite{Zhu2015b,Zhu2014} that the pairing between two holes can be significantly enhanced at $\alpha >\alpha_c$ beyond a simple RVB picture. In the following, an underlying mechanism is discussed based on the incoherent behavior of a single doped hole found in this paper at $\alpha >\alpha_c$.
 
According to the single-hole wavefunction in Eq.~\eqref{eqn:singleholeansatz}, the two-hole ground state may be constructed accordingly in the following form \cite{Zhao2022}: 
\begin{equation}
    |\Psi\rangle_{\mathrm{2h}} = \sum_{ij}g(i,j)c_{i\uparrow}c_{j\downarrow}e^{-i(\hat{\Omega}_i-\hat{\Omega}_j)}
    |\phi_0\rangle+\cdots~,
    \label{eqn:twoholeansz}
\end{equation}
where the two-hole wavefunction $g(i,j)$ will be taken as the variational parameter and the $\cdots$ term denotes the opposite chirality (the complex conjugate) of the phase-shift operator $e^{-i(\hat{\Omega}_i-\hat{\Omega}_j)}$. Such a two-hole state has been already variationally studied in Ref.~\onlinecite{Zhao2022} for both the 2D square lattice and the isotropic two-leg ladder and an extremely anisotropic two-leg ladder in Ref.~\onlinecite{Chen2018}. The results are in good and systematic agreement with the exact numerics. 

Our variational study of the wavefunction in Eq.~\eqref{eqn:twoholeansz} confirms the strong binding between the two holes at $\alpha>\alpha_c$. In particular, for each term in Eq.~\eqref{eqn:twoholeansz}, the distribution of the spin currents around the holes is shown in Fig.~\ref{fig:spincurrent2h} at different hole configurations (an opposite chirality is not shown) at $\alpha=1.0$. It shows that as two holes are separated spatially, opposite chiral spin currents emerge around the individual holes similar to the pattern in Fig.~\ref{fig:spincurrent}. Note that since the opposite chirality of the spin current in Fig. \ref{fig:spincurrent} corresponds to the opposite momentum $k^{\pm}_0$, such a pairing state is Cooper-pair-like with a zero total momentum and zero total spin. Once two holes are tightly paired in Fig.~\ref{fig:spincurrent2h}, the spin currents are completely canceled out.  

Such a spin current pattern for two holes is quite similar to the 2D case \cite{Zhao2022} where a roton-like (vortex-antivortex) pattern of the spin current has been identified as mediating the important pairing force between the holes. On the other hand, at $\alpha < \alpha_c$, the spin current is absent in the single-hole-doped case, where the hole simply behaves like a Landau quasiparticle with diminishing pairing strength \cite{Zhu2015b}. Therefore, the spin current associated with a non-Landau quasiparticle at $\alpha > \alpha_c$ is important to the pairing between the holes. In other words, the incoherent motion of the doped hole becomes the critical source for pairing by which the dynamic frustration as an intrinsic component of the single hole's motion can get eliminated. It is pointed out that by forming a tight pair, a charge $2e$ of two holes can be recovered \cite{Zhu2015b,Zhu2014}, which otherwise is diminished to zero in the unpaired single-hole case as studied in this paper. 


\section{Conclusion}\label{sec:conclusion}

The single-hole problem of a $t$-$J$ ladder is one of the simplest of doped Mott insulators, given the gapped (undoped) spin background for an even-leg ladder. Nevertheless, a novel phenomenon emerges in such a quasi-1D system, which can be entirely attributed to a quantum entanglement between the doped hole with the spin background via the nonlocal phase-string factor $e^{- i\hat{\Omega}_i}$ in the ground-state ansatz of Eq.~\eqref{eqn:1hgs}. It is nonperturbative in nature, whose effect is thoroughly explored by the VMC calculation in comparison with the exact numerics in this paper. 

Generally speaking, the phase-string effect incorporated by  $e^{- i\hat{\Omega}_i}$ in the single-hole wavefunction of Eq.~\eqref{eqn:1hgs} is important to facilitate the hopping of the doped hole on a (short-range) AF spin background as illustrated by the ground-state energies in Table~\ref{tab:relaxlad}. Since each hole is always accompanied by a spin-1/2 from a broken spin singlet pair, the effect of $e^{- i\hat{\Omega}_i}$ can further get explicitly exhibited via the behavior of such a hole-spin composite, whole size depends on the spin-spin correlation in the background. The detailed consequences as determined by the present VMC study are as follows.

Such a single-hole wavefunction ansatz can produce an excellent description of the phase diagram for the anisotropic two-leg $t$-$J$ ladder. It covers two distinct regions at $\alpha<\alpha_c$ and $\alpha>\alpha_c$, respectively. At $\alpha<\alpha_c$, it is a conventional Landau-like quasiparticle which can be adiabatically connected to a Bloch-wave of the bare hole created in the spin background $|\phi_0\rangle$:
\begin{equation}\label{Lqp} 
    c_{k_0\downarrow}|\phi_0\rangle
\end{equation}
at a non-degenerate $k_0=\pi \mod 2\pi$ with a total spin $S^z=1/2$. On the other hand, at $\alpha>\alpha_c$, a ``twisted'' quasiparticle emerges as
\begin{equation}\label{nqp}
    \tilde{c} _{k_0^\pm\downarrow}|\phi_0\rangle
\end{equation}
with double-degenerate ground state at $k_0^\pm=\pi \pm \kappa \mod 2\pi$ ($\kappa \neq 0$). Here $\tilde{c} _{k_0\downarrow}$ may be regarded as a 1D Bloch-wave (the Fourier transformation along the chain direction) of a twisted particle created by $\tilde{c} _{i\downarrow}\equiv {c}_{i\downarrow}e^{-i\hat{\Omega}_i}$. Since the half-filling ground state $|\phi_0\rangle$ is gapped, the usual longitudinal spin-polaron correction to either Eq.~\eqref{Lqp} or Eq.~\eqref{nqp} is perturbatively weak, which has been incorporated by the Lanczos method via $|\phi_0\rangle\rightarrow \hat{\Pi}_i|\phi_0\rangle$ with only quantitative improvements of the variational results.

It is found that at $\alpha>\alpha_c$, Eq.~\eqref{nqp} still has a finite overlap with Eq.~\eqref{Lqp} with the quasiparticle spectral weight $Z_{k_0^\pm}\neq 0$. But it does not mean that the two states can be smoothly connected to each other at the same $k_0^\pm$. The Landau's one-to-one correspondence principle is broken down here as there are two components in Eq.~\eqref{nqp} for a loosely bound hole-spin composite. Namely, besides a finite amplitude \cite{White2015} for a Landau quasiparticle, an incoherent component is also present in which a relative motion between the hole and spin-1/2 emerges inside the composite. In the latter, the relative spin current around the hole can carry away a continuum spectrum of momentum such that the hole becomes incoherent, which violates the \emph{charge} translational invariance \cite{Zhu2018a} and leads to a null response to an external magnetic flux. In fact, it is explicitly shown that the doped hole created by $\tilde{c} _{k_0^\pm\downarrow}$ becomes a charge-neutral spinon at $\alpha>\alpha_c$. The charge current carried by the doped hole disappears in the long ladder limit, while the spin current still remains unchanged as contributed by a spin $S^z=1/2$ at a finite momentum $k_0^\pm$. In contrast to the spin-charge separation in the 1D $t$-$J$ model chain, the charge of the doped hole simply ``vanishes'' in the two-leg ladder as the novel consequence of the \emph{transverse} spin-polaron effect introduced by $e^{- i\hat{\Omega}_i}$. 

In short, the earlier DMRG discovery \cite{Zhu2013} that the charge of the doped hole is self-localized while a neutral object (spinon) still behaves like a \emph{free} particle at $\alpha=1$ may be well reconciled with the latter DMRG result \cite{White2015} that there is no 1D-like spin-charge separation and the quasiparticle $Z_k\neq 0$. Based on the present wavefunction description, the doped hole actually becomes a loosely bound hole-spin composite here such that it still carries a well-defined spin-1/2 but its charge degree of freedom becomes incoherent due to the unscreened phase-string effect. As the essence of the doped Mott physics, the latter will generally break the charge translational symmetry, which may only be recovered either at $\alpha<\alpha_c$ or by pairing up of two holes \cite{Zhu2014,Zhao2022}. Lastly it is noted that the two-component structure in the present single-hole wavefunction may be further considered as a precursor of the ``Fermi arc'' physics at finite doping as recently explored in Ref.~\onlinecite{Zhang2022}. 

Finally, we have examined the pairing between two doped holes. As shown by DMRG \cite{Zhu2015b}, the pairing strength is rather weak at $\alpha<\alpha_c $, but gets substantially enhanced at $\alpha>\alpha_c $. It is due to the spin-current-carrying component that is strongly compensated once two holes form a tightly-bound pair to gain a substantial binding energy at $\alpha>\alpha_c $. It means that a novel pairing mechanism is also crucially related to the non-Landau behavior of the single hole at $\alpha>\alpha_c $ via the nonlocal phase-shift factor originated from the hidden phase-string sign structure in the $t$-$J$ model.    

\begin{acknowledgments}
Very helpful discussions with Qingrui Wang, Yang Qi, Zheng Zhu, Donna Sheng, Xiaoliang Qi, and Jan Zaanen are acknowledged. 
This work is partially supported by MOST of China (Grant No. 2021YFA1402101). 
\end{acknowledgments}

\appendix

\section{Phase-shift operator on a two-leg ladder with PBC} \label{app:1}


Despite that the essential physics should not be affected by boundary conditions in the long ladder limit, 
using PBC makes it more convenient to detect translationally invariant properties, such as momentum and any nonzero current configurations at finite-size calculations. 
It also makes it simpler to measure the response to external electromagnetic fields by inserting flux into the ring formed by the ladder. However, the original phase-shift operator $\hat{\Omega}_i$ is defined on a lattice under OBC \cite{Wang2015}. To capture these properties explicitly in the variational wavefunction Eq.~\eqref{eqn:singleholeansatz}, 
we need to generalize the original single-hole wavefunction \cite{Wang2015} to the case under PBC, 
\begin{equation}
    |\Psi_{\mathrm G}\rangle_{\mathrm{1h}} = \sum_{i,a} \varphi_h^{(a)}(i)e^{- i\hat{\Omega}_i^a}c_{i\bar{\sigma}}|\phi_0\rangle~,
    \label{eqn:singleholeansatzlad}
\end{equation}
where the phase-shift operator is defined by
\begin{equation}
    \hat{\Omega}^a_i = \sum_{l(\neq i)}\theta^a_i(l)n_{l\downarrow}~,
    \label{eqn:phasestringoperatorlad}
\end{equation}
and the definition of the statistical angles $\theta^a_i(l)$ on a PBC ladder is shown in Fig.~\ref{fig:PBCphase}. 
The two legs of the ladder are bent into two rings with different radii on the same plane, i.e.,  $r_{\mathrm{in}} = 2 - \lambda$ and $r_{\mathrm{out}} = 2 + \lambda$, 
with $\lambda$ a variational parameter to be determined by minimizing the variational energy. 
The phase $\theta^a_i(l)$ can be defined as a conventional 2D angle: $\theta^a_i(l) = \pm \mathrm{Im}\ln(z_i-z_l)$ 
where $z_i = x_i + i y_i$ is the complex coordinate of site $i$. The superscript $a = 1, 2$ here labels two choices of the radii for the two chains of the ladder, 
i.e. for $a = 1$ the first chain is bent into the inner ring and the second chain into the outer ring; 
while for $a = 2$ the second chain becomes the inner ring and the first chain the outer one. 

Under the PBC with explicit translational invariant symmetry, the variational parameter $\varphi_h^{(a)}$ is constraint to be a plane wave 
\begin{equation}
    \varphi_h^{(a)}(i) \propto e^{i\tilde{\mathbf{k}}^\pm\cdot \mathbf{r}_i}~, 
\end{equation}
where $\tilde{\mathbf{k}}^\pm = (\tilde{k}_x^\pm, \tilde{k}_y^\pm)$ is a variational parameter to be determined by minimizing the ground state energy. 
The relation between $\tilde{{k}}^\pm_x$ and the total momentum of the ground-state ${k}^\pm_x$ along the chain direction can be derived by 
a one-step translational transformation $\hat{T}$ along the chain direction acting on the translational invariant ground state $|\Psi_{\mathrm{G}}\rangle_{\mathrm{1h}}$
\begin{equation}
    \begin{aligned}
        \hat{T} |\Psi_{\mathrm G}\rangle_{\mathrm 1h} \equiv & e^{-ik_x^\pm} |\Psi_{\mathrm G}\rangle_{\mathrm 1h} \\
        = & \sum_i \varphi_h^{(a)}(i)e^{- i\sum_{l(\neq i)}\theta^a_i(l)n_{\mathcal{T}l,\downarrow}}c_{{\mathcal{T}i},\bar{\sigma}} |\phi_0\rangle\\
        =&\sum_i \varphi_h^{(a)}(\mathcal{T}^{-1}i)e^{- i\sum_{l(\neq i)}\theta_{\mathcal{T}^{-1}i}^a(\mathcal{T}^{-1}l)n_{l\downarrow}}c_{i\bar{\sigma}} |\phi_0\rangle\\
        =&e^{-i\tilde{k}_x^\pm } \sum_i \varphi_h^{(a)}(i) e^{- i \sum_{l(\neq i)} (\theta_i^a(l)\mp 2\pi/N_x) n_{l\downarrow} } c_{i\bar{\sigma}}|\phi_0\rangle\\
        =&e^{-i\tilde{k}_x^\pm \mp i (\sigma + 1) \pi/N_x }|\Psi_{\mathrm{G}}\rangle_{\mathrm 1h}~,
    \end{aligned}
\end{equation}
where we use $\mathcal{T}i$ and $\mathcal{T}^{-1}i$ to represent the site obtained by translating site $i$ one lattice constant along the positive and negative chain direction. 
The relation $\theta^a_{\mathcal{T}^{-1}i}(\mathcal{T}^{-1}l) = \theta^a_i(l) \mp 2\pi/N_x$ is used in deriving the fourth line of the equation.
The above expression shows that $k_x^\pm = \tilde{k}_x^\pm + K_x^\pm(\hat{\Omega})$, 
Where $K_x^\pm(\hat{\Omega}) = \pm(\sigma + 1) \pi / N_x$ denotes the momentum contributed by the phase factor $e^{\mp i\hat{\Omega}_i}$. 
As $K_x$ vanishes for $\sigma = -1$, and can be canceled for $\sigma = +1$ by a redefinition of the phase factor $\hat{\Omega}_i\rightarrow \sum_{l(\neq i)}\theta^a_i(l)n_{l\uparrow}$, we no longer distinguish $k_x$ and $\tilde{k}_x$ by only considering the $\sigma = -1$ case without loss of generality. 

For the $y$ component momentum ${k}_y$, we can get a similar expression $k_y^\pm = \tilde{k}_y^\pm + K_y^\pm(\hat{\Omega})$. 
But for a given index $a$, the $Z_2$ symmetry by exchanging the two chains is broken in a single configuration shown in Fig.~\ref{fig:PBCphase}. 
This symmetry can be restored after the summation over the index $a$, and the total momentum $k_y^\pm$ remains a good quantum number. 
For all the parameter regions we are interested in this article, we always find $k_y^\pm=0$ for the ground state. 
We therefore keep only the $x$ component momentum $k_x$ in the main text and denote it simply as $k$, 
keeping in mind that it represents a momentum along the chain direction with $k_y=0$. 

The advantage of the circular configuration shown in Fig.~\ref{fig:PBCphase} can be explicitly seen by taking the single-hole doped 1D $t$-$J$ chain as an example, 
which is detailed studied in Ref.~\onlinecite{Zhu2016}. 
By constructing a similar circular configuration to define $\theta_i(l)$ in a 1D ring, one sees that every time when the hole hops from site $i$ to its neighboring site $i\pm 1$, it will pick up a phase shift 
$\theta_i(l) - \theta_{i\pm 1}(l) = \pm \pi/N_x$ from each down spin at site $l$ (which is just the circular angle and thus independent to $i$ and $l$). 
When the phase shifts from all the down spins are summed over, one gets an effective phase 
$\pm N_\downarrow\pi/N_x\approx \pm \pi/2$ every time when the doped hole hops. 
Such $\pm \pi/2$ phase as contributed by all the background spins just constitutes the ground state momentum of a single hole doped $t$-$J$ chain, giving rise to a simple example of Anderson's phase shift idea. 

\begin{figure}
    \centering
    \includegraphics[width=0.46\textwidth]{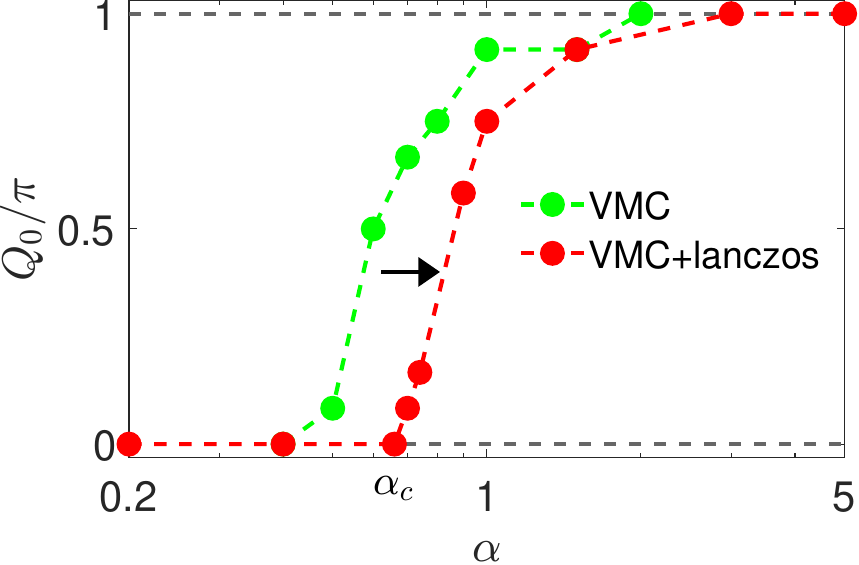}
    \caption{The phase diagrams based on $Q_0$ calculated by VMC before (green) and after (red) a power 1 Lanczos procedure is applied to incorporate a longitudinal spin-polaron effect. Note that the two phase diagrams are in qualitative agreement with only a global shift along the $\alpha$ axis with the correction. }
    \label{fig:alphaclanczos}
\end{figure}

\section{Longitudinal spin-polaron effect on the variational ansatz}\label{app:2}

In the main text, we have shown that the phase-shift operator $e^{- i\hat{\Omega}_i}$ plays the essential role 
in the non-Landau-type single-hole state of Eq.~\eqref{eqn:singleholeansatz}. It represents the singular phase-string effect, which results in the transverse spin current surrounding the doped hole to facilitate its hopping. The novel physical properties of the non-Landau quasiparticle at $\alpha>\alpha_c$ can be entirely attributed to such a transverse spin distortion as a many-body ``cloud'' associated with the hole. 

On the other hand, the conventional spin distortion or the ``longitudinal spin-polaron effect'' should be generally present in both regimes of $\alpha<\alpha_c$ and $\alpha>\alpha_c$. Since the spin background is gapped at half-filling, such a local spin distortion is expected to be perturbatively weak such that one may use an approximation similar to the power 1 Lanczos method to incorporate it to further improve the variational ground state $|\Psi_{\mathrm G}\rangle$. In the following we outline the corresponding procedure. 

For a trial state $|\Psi_{\mathrm G}\rangle$, the power 1 Lanczos method is to construct a new trial state $|\Psi_{\mathrm G}\rangle + \gamma H|\Psi_{\mathrm G}\rangle$. Here $H$ is the system Hamiltonian, and $\gamma$ is a new variational parameter to be optimized. Physically, such a one-step evolution of the Hamiltonian $H$ on $|\Psi_{\mathrm G}\rangle$ takes a local distortion into consideration, and the exact ground state can be achieved if the above procedure is iterated continuously until convergence. 

In the spirit of power 1 Lanczos method, then we consider $c_{i\downarrow}|\phi_0\rangle\rightarrow c_{i\downarrow}\hat{\Pi}_i|\phi_0\rangle$ with $\hat{\Pi}_i$ incorporating the local hopping effect produced by $H_t$ to the first order of correction, 
\begin{equation}\label{eqn:polaron} 
    c_{i\downarrow}\hat{\Pi}_i= c_{i\downarrow}+ \sum_{j\in \mathrm{NN}(i)} \left [a^1(i,j)c_{i\downarrow}n_{j\downarrow}+a^2(i,j)c_{i\uparrow} S_j^+\right]
\end{equation}
where $a^1(i,j)$ and $a^2(i,j)$ are two new variational parameters. Then a new variational wavefunction is constructed 
\begin{equation}\label{eqn:singleholeansatzlanczos} 
    |\tilde{\Psi}_{\mathrm G}\rangle _{\mathrm{1h}} = \sum_i \varphi^{(a)}_h(i)e^{- i\hat{\Omega}^a_i}c_{i\downarrow}\hat{\Pi}_i|\phi_0\rangle.
\end{equation}




Figure.~\ref{fig:alphaclanczos} shows the phase diagram determined by VMC before and after the above power 1 Lanczos method is applied. It shows that the overall line shape of $Q_0$ vs. $\alpha$ is unchanged, but there is a global shift along the $\alpha$ axis, which makes the VMC result in excellent agreement with the DMGR result as shown in Fig.~\ref{fig:kalpha}(b). 
We therefore conclude that although the qualitative physics are unchanged by the longitudinal local spin-polaron effect, which is solely decided by the transverse phase-shift field due to the phase-string, 
the local spin distortion can still effectively improve the variational energy, momentum $k_0^\pm$, and even the value of $\alpha_c$ to be in quantitative agreement with the DMRG results. 

Finally, we mention that the standard power Lanczos method up to power 2 is used in the benchmarking calculation of the ground-state energy in a small lattice size as shown in Table.~\ref{tab:relaxlad}, 
with the optimized wavefunction also denoted as $|\tilde{\Psi}_{\mathrm G}\rangle $ and $|\tilde{\Psi}_{\mathrm B}\rangle$. 
All the other VMC results presented in the main text are optimized based on Eq.~\eqref{eqn:singleholeansatzlanczos}. 

\section{Variational Monte Carlo procedure}\label{app:3}


The procedures of the variational Monte Carlo calculations were already explained in great detail in previous papers \cite{Wang2015,Chen2019,Chen2018,Zhao2022}. 
For completeness of the article, in this appendix, we briefly introduce the variational Monte Carlo procedures used 
to optimize the variational ground state and to measure other observables. 

\subsection{Variational procedure for optimizing ground state energy}
There are total two sets of variational parameters to be optimized, i.e., the anisotropic phase parameter $\lambda$ in the definition of $\hat{\Omega}$ and the variational wavefunction $\varphi_h (i)$. 
To optimize the ground-state energy, we first fix the value of $\lambda$, and calculate the ground-state energy as a function of $\varphi_h(i)$, 
which turns out to be a quadratic form as shown later in Eq.~\eqref{eqn:generalizeddiag_E}. 
The procedure to optimize the energy $E_{\mathrm{G}}$ with fixed $\lambda$ then turns out to be a generalized eigenvalue problem, which can be done by standard mathematic packages. 
Next, we plot the ground-state energy $E_{\mathrm{G}}$ as a function of different $\lambda$, and fit it with spline functions. 
The minimized energy and the corresponding $\lambda$ are then fixed by the minimal point of this fitted curve. 

\subsection{Monte Carlo procedure to calculate physical observables}

Before turning to the hole doped problem, we first need a wavefunction $|\phi_0\rangle$ of the half-filling state, 
where the $t$-$J$ model reduces to the Heisenberg spin model on a bipartite square lattice. 
It was shown that the ground state $|\phi_0\rangle$ can be well simulated by the Liang-Doucot-Anderson type wavefunction \cite{Liang1988}, 
\begin{equation} \label{eqn:Liang}
    |\phi_0\rangle = \sum_v\omega_v|v\rangle ~,
\end{equation}
where the valence bond (VB) state 
\begin{equation}
    |v\rangle = |(a_1,b_1)\cdots(a_n,b_n)\rangle~ 
\end{equation}
consist of singlet pairs $|(a,b)\rangle = |\uparrow_a\downarrow_b\rangle-|\downarrow_a\uparrow_b\rangle$. 
Here $a$ and $b$ comes from the two different sublattices A and B of the bipartite lattice, respectively. 
According to the Marshall sign rule on a bipartite lattice, 
the variational parameter $w_v$ of each valence bond state $|v\rangle$ is always positive, 
which can therefore be used to construct a probability distribution for the Monte Carlo procedure 
\begin{equation} \label{eqn:probabilitydistribution}
    P(v',v) = \frac{w_{v'}w_v\langle v'|v\rangle}{\langle\phi_0|\phi_0\rangle}~, 
\end{equation}
which satisfies $\sum_{v',v}P(v',v)=1$ and $P(v',v)>0$. 
The expectation value of any observable $\hat{O}$ can then be evaluated as 
\begin{equation} \label{eqn:expHF}
    \frac{\langle \phi_0| \hat{O} |\phi_0\rangle}{\langle \phi_0|\phi_0\rangle } = \sum_{v',v}P(v',v)O(v',v)~,
\end{equation}
as long as one gets the expression of $O(v',v)$. 

After a hole is doped into the spin background $|\phi_0\rangle$, 
we follow the same idea to express all the observables on $|\Psi_{\mathrm G}\rangle$ in terms of expectation values of the spin background $|\phi_0\rangle$. 
By fixing the normalization condition of $|\Psi_{\mathrm{G}}\rangle_{\mathrm{1h}}$ as 
\begin{equation}
    \frac{\tensor[_{1\mathrm{h}}]{\langle \Psi_{\mathrm{G}}|\hat{\mathds{1}}|\Psi_{\mathrm{G}}\rangle}{_{1\mathrm{h}}}}{\langle \phi_0|\phi_0\rangle} = 1~,
    \label{eqn:norm}
\end{equation}
the expectation value of an arbitrary observable $\hat{O}$ is then a quadratic form of the variational parameters $\varphi^a_h(i)$ 
\begin{equation}
    \langle\hat{O}\rangle \equiv 
    \frac{\tensor[_{1\mathrm{h}}]{\langle \Psi_{\mathrm{G}}|\hat{O}|\Psi_{\mathrm{G}}\rangle}{_{1\mathrm{h}}}}{\langle \phi_0|\phi_0\rangle}
    =\boldsymbol{\varphi}^\dagger \mathbf{O}\boldsymbol{\varphi}~,
    \label{eqn:opeO}
\end{equation}
where the variational parameters $\varphi_h^a(i)$ are written as a vector $\boldsymbol{\varphi}$, 
and $\mathbf{O}$ is a Hermitian matrix with its matrix elements as 
\begin{equation} \label{eqn:matrixO}
    \mathbf{O}_{ia}^{i'a'} = \sum_{v',v}P(v',v)\frac{\langle v'|e^{+ i\hat{\Omega}^{a'}_{i'}} c^{\dagger}_{i'\downarrow}\hat{O} 
        c^{}_{i\downarrow}e^{- i\hat{\Omega}_{i}^{a}}|v\rangle} {\langle v'|v\rangle}~. 
\end{equation}
The Monte Carlo sampling procedure can then be used here to measure the matrix elements in Eq.~\eqref{eqn:matrixO} 
by sampling different transposition-graph covers $(v',v)$. 
To improve the Monte Carlo sampling efficiency, the loop update procedure introduced in Ref.~\cite{Sandvik2010} is used here in the updating progress of $(v',v)$. 

We note that Eq.~\eqref{eqn:norm} actually also falls into the form of Eq.~\eqref{eqn:opeO}, 
with $\hat{O}$ taken as the identity operator $\hat{\mathds{1}}$, 
\begin{equation} \label{eqn:ope1}
    1 = \langle \hat{\mathds{1}}\rangle =\boldsymbol{\varphi}^\dagger \mathbf{A}\boldsymbol{\varphi}~,
\end{equation}
which is also a quadratic form with the corresponding matrix denoted as $\mathbf{A}$ here. 
To get the variational energies, we take the operator $\hat{O}$ in Eq.~\eqref{eqn:opeO} as the Hamiltonian $\hat{H}$ 
\begin{equation} 
    E_{\mathrm G} \equiv \langle \hat{H}\rangle = \boldsymbol{\varphi}^\dagger \mathbf{H} \boldsymbol{\varphi} \label{eqn:generalizeddiag_E}~.
\end{equation}
Combined with the normalization condition Eq.~\eqref{eqn:ope1}, 
the procedure to optimize $E_{\mathrm G}$ then turns out to be a generalized eigenvalue problem 
\begin{equation}\label{eqn:generalizeddiag}
    \mathbf{H}\boldsymbol{\varphi} = E_{\mathrm G} \mathbf{A}\boldsymbol{\varphi}~.
\end{equation}

To explicitly calculate the matrix elements Eq.~\eqref{eqn:matrixO}, we follow Ref.~\onlinecite{Zhao2022}
to transform all the operators to the following form: 
\begin{equation}
    \frac{\langle v'|e^{+ i\hat{\Omega}_{i'}^{a'}} n_{k_1\sigma_1}n_{k_2\sigma_2}\cdots n_{k_n\sigma_n}
    S_{l_1}^{\sigma_1}S_{l_2}^{\sigma_2}\cdots S_{l_s}^{\sigma_s}
    e^{- i\hat{\Omega}_i^a}|v\rangle }{\langle v'|v\rangle }~.
    \label{eqn:nnnsss}
\end{equation}
Then for every pair of $v'$ and $v$, one can check the compatibility of the $n_{k_i\sigma_i}$ s and $S_{l_j}^{\sigma_j}$ s with the 
loop configurations in the transposition-graph covers $(v',v)$, as well as calculate the phase factors $e^{i\hat{\Omega}_i^a}$ for every loop. 
In the following we give the explicit formulas of Eq.~\eqref{eqn:matrixO} for some important operators measured in the main text.

For the normalization condition where $\hat{O}=\hat{\mathds{1}}$, the matrix elements of $\mathbf{A}$ are given by 
\begin{equation}
    \mathbf{A}^{i'a'}_{ia}= \sum_{v',v}P(v',v)\frac{1}{\langle v'|v\rangle}\delta_{ii'}
    \langle v'|e^{+ i\hat{\Omega}_{i'}^{a'}}n_{i\downarrow}e^{- i\hat{\Omega}_i^a}|v\rangle ~.
    \label{eqn:measurenorm}
\end{equation}
The hole density at every site $n_i^h$ can be directed evaluated from the $\mathbf{A}$ matrix with 
\begin{equation}
    n^h_i = \sum_{a,a'}\mathbf{A}_{ia}^{ia'}~.
\end{equation}

The hopping Hamiltonian $\hat{H}_t$ Eq.~\eqref{eqn:hoppingH} connects two NN sites $i$ and $i'\pm \mathbf{e}_\alpha$, 
\begin{equation}
    \begin{aligned}
        (\mathbf{H}_{t})_{ia}^{i'a'} = &t\sum_{v',v,\alpha = x,y}P(v',v)\frac{1}{\langle v'|v\rangle}
        \delta_{ii'\pm\mathbf{e}_\alpha}\\
        &\times \langle v'|e^{+ i\hat{\Omega}_{i'}^{a'}}(n_{i\downarrow}n_{i'\downarrow}
        +S_i^+S_{i'}^-)e^{- i\hat{\Omega}_i^a}|v\rangle ~,
    \end{aligned}
    \label{eqn:measurehopping}
\end{equation}
where $\mathbf{e}_{x,y}$ are the $x$ and $y$ direction unit vectors. 
The charge current $J^c_{ij}$ can also be derived from the hopping matrix by just taking the imaginary part of it 
\begin{equation}
    J^c_{ij} = 2\sum_{a,a'}\mathrm{Im}\left((\varphi^{a'}_h(j))^*(\mathbf{H}_t)_{ia}^{ja'}\varphi_h^a(i)\right)~.
\end{equation}

The superexchange Hamiltonian $\hat{H}_J$ Eq.~\eqref{eqn:superexchangeH} does not change the position of the hole 
\begin{equation}
    \begin{aligned}
        (\mathbf{H}_{J})_{ia}^{i'a'} = &\frac{J}{2} \sum_{v',v,\langle kl\rangle(\neq i)}P(v',v)\frac{1}{\langle v'|v\rangle} \delta_{ii'}
        \langle v'|e^{+ i\hat{\Omega}_{i'}^{a'}}n_{i\downarrow}\\ &\times
        (S_k^+S_l^-+S_k^-S_l^+-n_{k\uparrow}n_{l\downarrow}-n_{k\downarrow}n_{l\uparrow})e^{- i\hat{\Omega}_i^a}|v\rangle 
    \end{aligned}
    \label{eqn:measureAF}
\end{equation}

The neutral spin current $\hat{J}_{kl}^s$ defined in Eq.~\eqref{eqn:spincurrent1} has similar matrix elements with the superexchange term, 
\begin{equation}
    \begin{aligned}
        (\mathbf{J}_{kl}^s)^{i'a'}_{ia} =& -i\frac{J}{2} (1-\delta_{ik})(1-\delta_{il})\sum_{v',v}P(v',v)\frac{1}{\langle v'|v\rangle} \delta_{ii'}\\
        &\times \langle v'|e^{+ i\hat{\Omega}_{i'}^{a'}}n_{i\downarrow} (S_k^+S_l^--S_k^-S_l^+)e^{- i\hat{\Omega}_i^a}|v\rangle. 
    \end{aligned}
    \label{eqn:measureJs}
\end{equation}

Finally, for the quasiparticle weight, $Z_k$, we need to normalize Eq.~\eqref{eqn:Zkdefin} as 
\begin{equation}
    \begin{aligned}
        Z_{k} =& \left| \frac{\langle\phi_0|c_{k\downarrow}^\dagger|\Psi_{\mathrm G}\rangle_{\mathrm 1h}}
        {\sqrt{\langle\phi_0|\phi_0\rangle\tensor[_{\mathrm 1h}]{\langle\Psi_{\mathrm G}|\Psi_{\mathrm G}\rangle}{_{\mathrm 1h}}}} \right|^2\\
        =&\left|\sum_{i,a}\varphi^a_h(i)\frac{ \langle\phi_0|c^\dagger_{k\downarrow}c^{}_{i\downarrow}e^{- i\hat{\Omega}^{a}_{i}}|\phi_0\rangle}{\langle\phi_0|\phi_0\rangle} \right|^2 \\
        =&\left|\sum_{i'a',ia}\frac{1}{N}e^{ikx_{i'}}\varphi^a_h(i)\sum_{v',v}P(v',v)
        \frac{\langle v'|n_{i\downarrow}e^{- i\hat{\Omega}_i^a}|v\rangle} {\langle v'|v\rangle}\right|^2,\\
    \end{aligned}
\end{equation}
which is a similar quadratic formula with Eq.~\eqref{eqn:opeO} if we treat $\frac{1}{N}e^{-ikx_{i'}}$ as 
a new wavefunction with $a=0$ as $\varphi^0_h(i')$. 

When the Lanczos procedure in Appendix~\ref{app:2} is taken into considered, the above expressions will become too complicated to be calculated by hands. 
Actually, we write a program to generate the expressions automatically in realistic calculations, and the explicit expressions will be not given here.


\bibliography{ref/nameAbrv.bib,ref/refs.bib}

\begin{thebibliography}{31}%
\makeatletter
\providecommand \@ifxundefined [1]{%
 \@ifx{#1\undefined}
}%
\providecommand \@ifnum [1]{%
 \ifnum #1\expandafter \@firstoftwo
 \else \expandafter \@secondoftwo
 \fi
}%
\providecommand \@ifx [1]{%
 \ifx #1\expandafter \@firstoftwo
 \else \expandafter \@secondoftwo
 \fi
}%
\providecommand \natexlab [1]{#1}%
\providecommand \enquote  [1]{``#1''}%
\providecommand \bibnamefont  [1]{#1}%
\providecommand \bibfnamefont [1]{#1}%
\providecommand \citenamefont [1]{#1}%
\providecommand \href@noop [0]{\@secondoftwo}%
\providecommand \href [0]{\begingroup \@sanitize@url \@href}%
\providecommand \@href[1]{\@@startlink{#1}\@@href}%
\providecommand \@@href[1]{\endgroup#1\@@endlink}%
\providecommand \@sanitize@url [0]{\catcode `\\12\catcode `\$12\catcode
  `\&12\catcode `\#12\catcode `\^12\catcode `\_12\catcode `\%12\relax}%
\providecommand \@@startlink[1]{}%
\providecommand \@@endlink[0]{}%
\providecommand \url  [0]{\begingroup\@sanitize@url \@url }%
\providecommand \@url [1]{\endgroup\@href {#1}{\urlprefix }}%
\providecommand \urlprefix  [0]{URL }%
\providecommand \Eprint [0]{\href }%
\providecommand \doibase [0]{http://dx.doi.org/}%
\providecommand \selectlanguage [0]{\@gobble}%
\providecommand \bibinfo  [0]{\@secondoftwo}%
\providecommand \bibfield  [0]{\@secondoftwo}%
\providecommand \translation [1]{[#1]}%
\providecommand \BibitemOpen [0]{}%
\providecommand \bibitemStop [0]{}%
\providecommand \bibitemNoStop [0]{.\EOS\space}%
\providecommand \EOS [0]{\spacefactor3000\relax}%
\providecommand \BibitemShut  [1]{\csname bibitem#1\endcsname}%
\let\auto@bib@innerbib\@empty
\bibitem [{\citenamefont {Mott}(1949)}]{Mott1949}%
  \BibitemOpen
  \bibfield  {author} {\bibinfo {author} {\bibfnamefont {N.~F.}\ \bibnamefont
  {Mott}},\ }\bibfield  {title} {\enquote {\bibinfo {title} {{The Basis of the
  Electron Theory of Metals, with Special Reference to the Transition
  Metals}},}\ }\href {\doibase 10.1088/0370-1298/62/7/303} {\bibfield
  {journal} {\bibinfo  {journal} {Proc. Phys. Soc. A}\ }\textbf {\bibinfo
  {volume} {62}},\ \bibinfo {pages} {416--422} (\bibinfo {year}
  {1949})}\BibitemShut {NoStop}%
\bibitem [{\citenamefont {Anderson}(1987)}]{Anderson1987}%
  \BibitemOpen
  \bibfield  {author} {\bibinfo {author} {\bibfnamefont {P.~W.}\ \bibnamefont
  {Anderson}},\ }\bibfield  {title} {\enquote {\bibinfo {title} {{The
  Resonating Valence Bond State in $\mathrm{La}_2\mathrm{CuO}_4$ and
  Superconductivity}},}\ }\href {\doibase 10.1126/science.235.4793.1196}
  {\bibfield  {journal} {\bibinfo  {journal} {Science}\ }\textbf {\bibinfo
  {volume} {235}},\ \bibinfo {pages} {1196--1198} (\bibinfo {year}
  {1987})}\BibitemShut {NoStop}%
\bibitem [{\citenamefont {Imada}\ \emph {et~al.}(1998)\citenamefont {Imada},
  \citenamefont {Fujimori},\ and\ \citenamefont {Tokura}}]{Imada1998}%
  \BibitemOpen
  \bibfield  {author} {\bibinfo {author} {\bibfnamefont {Masatoshi}\
  \bibnamefont {Imada}}, \bibinfo {author} {\bibfnamefont {Atsushi}\
  \bibnamefont {Fujimori}}, \ and\ \bibinfo {author} {\bibfnamefont
  {Yoshinori}\ \bibnamefont {Tokura}},\ }\bibfield  {title} {\enquote {\bibinfo
  {title} {{Metal-Insulator Transitions}},}\ }\href {\doibase
  10.1103/RevModPhys.70.1039} {\bibfield  {journal} {\bibinfo  {journal} {Rev.
  Mod. Phys.}\ }\textbf {\bibinfo {volume} {70}},\ \bibinfo {pages}
  {1039--1263} (\bibinfo {year} {1998})}\BibitemShut {NoStop}%
\bibitem [{\citenamefont {Lee}\ \emph {et~al.}(2006)\citenamefont {Lee},
  \citenamefont {Nagaosa},\ and\ \citenamefont {Wen}}]{Lee2006}%
  \BibitemOpen
  \bibfield  {author} {\bibinfo {author} {\bibfnamefont {Patrick~A.}\
  \bibnamefont {Lee}}, \bibinfo {author} {\bibfnamefont {Naoto}\ \bibnamefont
  {Nagaosa}}, \ and\ \bibinfo {author} {\bibfnamefont {Xiao~Gang}\ \bibnamefont
  {Wen}},\ }\bibfield  {title} {\enquote {\bibinfo {title} {{Doping a Mott
  Insulator: Physics of High-Temperature Superconductivity}},}\ }\href
  {\doibase 10.1103/RevModPhys.78.17} {\bibfield  {journal} {\bibinfo
  {journal} {Rev. Mod. Phys.}\ }\textbf {\bibinfo {volume} {78}},\ \bibinfo
  {pages} {17--85} (\bibinfo {year} {2006})},\ \Eprint
  {http://arxiv.org/abs/cond-mat/0410445} {arXiv:cond-mat/0410445} \BibitemShut
  {NoStop}%
\bibitem [{\citenamefont {Schmitt-Rink}\ \emph {et~al.}(1988)\citenamefont
  {Schmitt-Rink}, \citenamefont {Varma},\ and\ \citenamefont
  {Ruckenstein}}]{Schmitt-Rink1988}%
  \BibitemOpen
  \bibfield  {author} {\bibinfo {author} {\bibfnamefont {S.}~\bibnamefont
  {Schmitt-Rink}}, \bibinfo {author} {\bibfnamefont {C.~M.}\ \bibnamefont
  {Varma}}, \ and\ \bibinfo {author} {\bibfnamefont {A.~E.}\ \bibnamefont
  {Ruckenstein}},\ }\bibfield  {title} {\enquote {\bibinfo {title} {{Spectral
  Function of Holes in a Quantum Antiferromagnet}},}\ }\href {\doibase
  10.1103/PhysRevLett.60.2793} {\bibfield  {journal} {\bibinfo  {journal}
  {Phys. Rev. Lett.}\ }\textbf {\bibinfo {volume} {60}},\ \bibinfo {pages}
  {2793--2796} (\bibinfo {year} {1988})}\BibitemShut {NoStop}%
\bibitem [{\citenamefont {Kane}\ \emph {et~al.}(1989)\citenamefont {Kane},
  \citenamefont {Lee},\ and\ \citenamefont {Read}}]{Kane1989}%
  \BibitemOpen
  \bibfield  {author} {\bibinfo {author} {\bibfnamefont {C.~L.}\ \bibnamefont
  {Kane}}, \bibinfo {author} {\bibfnamefont {P.~A.}\ \bibnamefont {Lee}}, \
  and\ \bibinfo {author} {\bibfnamefont {N.}~\bibnamefont {Read}},\ }\bibfield
  {title} {\enquote {\bibinfo {title} {{Motion of a Single Hole in a Quantum
  Antiferromagnet}},}\ }\href {\doibase 10.1103/PhysRevB.39.6880} {\bibfield
  {journal} {\bibinfo  {journal} {Phys. Rev. B}\ }\textbf {\bibinfo {volume}
  {39}},\ \bibinfo {pages} {6880--6897} (\bibinfo {year} {1989})}\BibitemShut
  {NoStop}%
\bibitem [{\citenamefont {Anderson}(1990)}]{Anderson1990}%
  \BibitemOpen
  \bibfield  {author} {\bibinfo {author} {\bibfnamefont {P.~W.}\ \bibnamefont
  {Anderson}},\ }\bibfield  {title} {\enquote {\bibinfo {title}
  {{‘‘Luttinger-Liquid'' Behavior of the Normal Metallic State of the 2D
  Hubbard Model}},}\ }\href {\doibase 10.1103/PhysRevLett.64.1839} {\bibfield
  {journal} {\bibinfo  {journal} {Phys. Rev. Lett.}\ }\textbf {\bibinfo
  {volume} {64}},\ \bibinfo {pages} {1839--1841} (\bibinfo {year}
  {1990})}\BibitemShut {NoStop}%
\bibitem [{\citenamefont {Sheng}\ \emph {et~al.}(1996)\citenamefont {Sheng},
  \citenamefont {Chen},\ and\ \citenamefont {Weng}}]{Sheng1996}%
  \BibitemOpen
  \bibfield  {author} {\bibinfo {author} {\bibfnamefont {D.~N.}\ \bibnamefont
  {Sheng}}, \bibinfo {author} {\bibfnamefont {Y.~C.}\ \bibnamefont {Chen}}, \
  and\ \bibinfo {author} {\bibfnamefont {Z.~Y.}\ \bibnamefont {Weng}},\
  }\bibfield  {title} {\enquote {\bibinfo {title} {{Phase String Effect in a
  Doped Antiferromagnet}},}\ }\href {\doibase 10.1103/PhysRevLett.77.5102}
  {\bibfield  {journal} {\bibinfo  {journal} {Phys. Rev. Lett.}\ }\textbf
  {\bibinfo {volume} {77}},\ \bibinfo {pages} {5102--5105} (\bibinfo {year}
  {1996})}\BibitemShut {NoStop}%
\bibitem [{\citenamefont {Wu}\ \emph {et~al.}(2008)\citenamefont {Wu},
  \citenamefont {Weng},\ and\ \citenamefont {Zaanen}}]{Wu2008}%
  \BibitemOpen
  \bibfield  {author} {\bibinfo {author} {\bibfnamefont {K.}~\bibnamefont
  {Wu}}, \bibinfo {author} {\bibfnamefont {Z.~Y.}\ \bibnamefont {Weng}}, \ and\
  \bibinfo {author} {\bibfnamefont {J.}~\bibnamefont {Zaanen}},\ }\bibfield
  {title} {\enquote {\bibinfo {title} {{Sign Structure of the $t$-$J$
  model}},}\ }\href {\doibase 10.1103/PhysRevB.77.155102} {\bibfield  {journal}
  {\bibinfo  {journal} {Phys. Rev. B}\ }\textbf {\bibinfo {volume} {77}},\
  \bibinfo {pages} {155102} (\bibinfo {year} {2008})},\ \Eprint
  {http://arxiv.org/abs/0802.0273} {arXiv:0802.0273} \BibitemShut {NoStop}%
\bibitem [{\citenamefont {Zhang}\ and\ \citenamefont {Weng}(2014)}]{Zhang2014}%
  \BibitemOpen
  \bibfield  {author} {\bibinfo {author} {\bibfnamefont {Long}\ \bibnamefont
  {Zhang}}\ and\ \bibinfo {author} {\bibfnamefont {Zheng~Yu}\ \bibnamefont
  {Weng}},\ }\bibfield  {title} {\enquote {\bibinfo {title} {{Sign structure,
  electron fractionalization, and emergent gauge description of the Hubbard
  model}},}\ }\href {\doibase 10.1103/PhysRevB.90.165120} {\bibfield  {journal}
  {\bibinfo  {journal} {Phys. Rev. B}\ }\textbf {\bibinfo {volume} {90}},\
  \bibinfo {pages} {165120} (\bibinfo {year} {2014})},\ \Eprint
  {http://arxiv.org/abs/1406.6867} {arXiv:1406.6867} \BibitemShut {NoStop}%
\bibitem [{\citenamefont {Zheng}\ \emph {et~al.}(2018)\citenamefont {Zheng},
  \citenamefont {Zhu}, \citenamefont {Sheng},\ and\ \citenamefont
  {Weng}}]{Zheng2018b}%
  \BibitemOpen
  \bibfield  {author} {\bibinfo {author} {\bibfnamefont {Wayne}\ \bibnamefont
  {Zheng}}, \bibinfo {author} {\bibfnamefont {Zheng}\ \bibnamefont {Zhu}},
  \bibinfo {author} {\bibfnamefont {D.~N.}\ \bibnamefont {Sheng}}, \ and\
  \bibinfo {author} {\bibfnamefont {Zheng-Yu}\ \bibnamefont {Weng}},\
  }\bibfield  {title} {\enquote {\bibinfo {title} {{Hidden Spin Current in
  Doped Mott Antiferromagnets}},}\ }\href {\doibase 10.1103/PhysRevB.98.165102}
  {\bibfield  {journal} {\bibinfo  {journal} {Phys. Rev. B}\ }\textbf {\bibinfo
  {volume} {98}},\ \bibinfo {pages} {165102} (\bibinfo {year} {2018})},\
  \Eprint {http://arxiv.org/abs/1802.05977} {arXiv:1802.05977} \BibitemShut
  {NoStop}%
\bibitem [{\citenamefont {Chen}\ \emph {et~al.}(2019)\citenamefont {Chen},
  \citenamefont {Wang}, \citenamefont {Qi}, \citenamefont {Sheng},\ and\
  \citenamefont {Weng}}]{Chen2019}%
  \BibitemOpen
  \bibfield  {author} {\bibinfo {author} {\bibfnamefont {Shuai}\ \bibnamefont
  {Chen}}, \bibinfo {author} {\bibfnamefont {Qing-Rui}\ \bibnamefont {Wang}},
  \bibinfo {author} {\bibfnamefont {Yang}\ \bibnamefont {Qi}}, \bibinfo
  {author} {\bibfnamefont {D.~N.}\ \bibnamefont {Sheng}}, \ and\ \bibinfo
  {author} {\bibfnamefont {Zheng-Yu}\ \bibnamefont {Weng}},\ }\bibfield
  {title} {\enquote {\bibinfo {title} {{Single-Hole Wave Function in Two
  Dimensions: A Case Study of the Doped Mott Insulator}},}\ }\href {\doibase
  10.1103/PhysRevB.99.205128} {\bibfield  {journal} {\bibinfo  {journal} {Phys.
  Rev. B}\ }\textbf {\bibinfo {volume} {99}},\ \bibinfo {pages} {205128}
  (\bibinfo {year} {2019})},\ \Eprint {http://arxiv.org/abs/1812.05627}
  {arXiv:1812.05627} \BibitemShut {NoStop}%
\bibitem [{\citenamefont {Shraiman}\ and\ \citenamefont
  {Siggia}(1988)}]{Shraiman1988a}%
  \BibitemOpen
  \bibfield  {author} {\bibinfo {author} {\bibfnamefont {Boris~I.}\
  \bibnamefont {Shraiman}}\ and\ \bibinfo {author} {\bibfnamefont {Eric~D.}\
  \bibnamefont {Siggia}},\ }\bibfield  {title} {\enquote {\bibinfo {title}
  {{Mobile Vacancies in a Quantum Heisenberg Antiferromagnet}},}\ }\href
  {\doibase 10.1103/PhysRevLett.61.467} {\bibfield  {journal} {\bibinfo
  {journal} {Phys. Rev. Lett.}\ }\textbf {\bibinfo {volume} {61}},\ \bibinfo
  {pages} {467--470} (\bibinfo {year} {1988})}\BibitemShut {NoStop}%
\bibitem [{\citenamefont {Shraiman}\ and\ \citenamefont
  {Siggia}(1989)}]{Shraiman1989}%
  \BibitemOpen
  \bibfield  {author} {\bibinfo {author} {\bibfnamefont {Boris~I.}\
  \bibnamefont {Shraiman}}\ and\ \bibinfo {author} {\bibfnamefont {Eric~D.}\
  \bibnamefont {Siggia}},\ }\bibfield  {title} {\enquote {\bibinfo {title}
  {{Spiral Phase of a Doped Quantum Antiferromagnet}},}\ }\href {\doibase
  10.1103/PhysRevLett.62.1564} {\bibfield  {journal} {\bibinfo  {journal}
  {Phys. Rev. Lett.}\ }\textbf {\bibinfo {volume} {62}},\ \bibinfo {pages}
  {1564--1567} (\bibinfo {year} {1989})}\BibitemShut {NoStop}%
\bibitem [{\citenamefont {Weng}(1991)}]{Weng1991}%
  \BibitemOpen
  \bibfield  {author} {\bibinfo {author} {\bibfnamefont {Z.~Y.}\ \bibnamefont
  {Weng}},\ }\bibfield  {title} {\enquote {\bibinfo {title} {{Dynamical Spiral
  State at Finite Doping}},}\ }\href {\doibase 10.1103/PhysRevLett.66.2156}
  {\bibfield  {journal} {\bibinfo  {journal} {Phys. Rev. Lett.}\ }\textbf
  {\bibinfo {volume} {66}},\ \bibinfo {pages} {2156--2159} (\bibinfo {year}
  {1991})}\BibitemShut {NoStop}%
\bibitem [{\citenamefont {Zhu}\ \emph {et~al.}(2013)\citenamefont {Zhu},
  \citenamefont {Jiang}, \citenamefont {Qi}, \citenamefont {Tian},\ and\
  \citenamefont {Weng}}]{Zhu2013}%
  \BibitemOpen
  \bibfield  {author} {\bibinfo {author} {\bibfnamefont {Zheng}\ \bibnamefont
  {Zhu}}, \bibinfo {author} {\bibfnamefont {Hong-Chen}\ \bibnamefont {Jiang}},
  \bibinfo {author} {\bibfnamefont {Yang}\ \bibnamefont {Qi}}, \bibinfo
  {author} {\bibfnamefont {Chushun}\ \bibnamefont {Tian}}, \ and\ \bibinfo
  {author} {\bibfnamefont {Zheng-Yu}\ \bibnamefont {Weng}},\ }\bibfield
  {title} {\enquote {\bibinfo {title} {{Strong Correlation Induced Charge
  Localization in Antiferromagnets}},}\ }\href {\doibase 10.1038/srep02586}
  {\bibfield  {journal} {\bibinfo  {journal} {Solar Phys.}\ }\textbf {\bibinfo
  {volume} {3}},\ \bibinfo {pages} {2586} (\bibinfo {year} {2013})},\ \Eprint
  {http://arxiv.org/abs/1212.6634} {arXiv:1212.6634} \BibitemShut {NoStop}%
\bibitem [{\citenamefont {Zhu}\ and\ \citenamefont {Weng}(2015)}]{Zhu2015b}%
  \BibitemOpen
  \bibfield  {author} {\bibinfo {author} {\bibfnamefont {Zheng}\ \bibnamefont
  {Zhu}}\ and\ \bibinfo {author} {\bibfnamefont {Zheng-Yu}\ \bibnamefont
  {Weng}},\ }\bibfield  {title} {\enquote {\bibinfo {title} {{Quasiparticle
  Collapsing in an Anisotropic $t$-$J$ Ladder}},}\ }\href {\doibase
  10.1103/PhysRevB.92.235156} {\bibfield  {journal} {\bibinfo  {journal} {Phys.
  Rev. B}\ }\textbf {\bibinfo {volume} {92}},\ \bibinfo {pages} {235156}
  (\bibinfo {year} {2015})},\ \Eprint {http://arxiv.org/abs/1409.3241}
  {arXiv:1409.3241} \BibitemShut {NoStop}%
\bibitem [{\citenamefont {Zhu}\ \emph {et~al.}(2015)\citenamefont {Zhu},
  \citenamefont {Tian}, \citenamefont {Jiang}, \citenamefont {Qi},
  \citenamefont {Weng},\ and\ \citenamefont {Zaanen}}]{Zhu2015}%
  \BibitemOpen
  \bibfield  {author} {\bibinfo {author} {\bibfnamefont {Zheng}\ \bibnamefont
  {Zhu}}, \bibinfo {author} {\bibfnamefont {Chushun}\ \bibnamefont {Tian}},
  \bibinfo {author} {\bibfnamefont {Hong-Chen}\ \bibnamefont {Jiang}}, \bibinfo
  {author} {\bibfnamefont {Yang}\ \bibnamefont {Qi}}, \bibinfo {author}
  {\bibfnamefont {Zheng-Yu}\ \bibnamefont {Weng}}, \ and\ \bibinfo {author}
  {\bibfnamefont {Jan}\ \bibnamefont {Zaanen}},\ }\bibfield  {title} {\enquote
  {\bibinfo {title} {Charge modulation as fingerprints of phase-string
  triggered interference},}\ }\href {\doibase 10.1103/PhysRevB.92.035113}
  {\bibfield  {journal} {\bibinfo  {journal} {Phys. Rev. B}\ }\textbf {\bibinfo
  {volume} {92}},\ \bibinfo {pages} {035113} (\bibinfo {year}
  {2015})}\BibitemShut {NoStop}%
\bibitem [{\citenamefont {Zhu}\ \emph {et~al.}(2018)\citenamefont {Zhu},
  \citenamefont {Sheng},\ and\ \citenamefont {Weng}}]{Zhu2018a}%
  \BibitemOpen
  \bibfield  {author} {\bibinfo {author} {\bibfnamefont {Zheng}\ \bibnamefont
  {Zhu}}, \bibinfo {author} {\bibfnamefont {D.~N.}\ \bibnamefont {Sheng}}, \
  and\ \bibinfo {author} {\bibfnamefont {Zheng-Yu}\ \bibnamefont {Weng}},\
  }\bibfield  {title} {\enquote {\bibinfo {title} {{Intrinsic Translational
  Symmetry Breaking in a Doped Mott Insulator}},}\ }\href {\doibase
  10.1103/PhysRevB.98.035129} {\bibfield  {journal} {\bibinfo  {journal} {Phys.
  Rev. B}\ }\textbf {\bibinfo {volume} {98}},\ \bibinfo {pages} {035129}
  (\bibinfo {year} {2018})},\ \Eprint {http://arxiv.org/abs/1707.00068}
  {arXiv:1707.00068} \BibitemShut {NoStop}%
\bibitem [{\citenamefont {White}\ \emph {et~al.}(2015)\citenamefont {White},
  \citenamefont {Scalapino},\ and\ \citenamefont {Kivelson}}]{White2015}%
  \BibitemOpen
  \bibfield  {author} {\bibinfo {author} {\bibfnamefont {S.~R.}\ \bibnamefont
  {White}}, \bibinfo {author} {\bibfnamefont {D.~J.}\ \bibnamefont
  {Scalapino}}, \ and\ \bibinfo {author} {\bibfnamefont {S.~A.}\ \bibnamefont
  {Kivelson}},\ }\bibfield  {title} {\enquote {\bibinfo {title} {{One Hole in
  the Two-Leg $t$-$J$ Ladder and Adiabatic Continuity to the Noninteracting
  Limit}},}\ }\href {\doibase 10.1103/PhysRevLett.115.056401} {\bibfield
  {journal} {\bibinfo  {journal} {Phys. Rev. Lett.}\ }\textbf {\bibinfo
  {volume} {115}},\ \bibinfo {pages} {056401} (\bibinfo {year} {2015})},\
  \Eprint {http://arxiv.org/abs/1502.04403} {arXiv:1502.04403} \BibitemShut
  {NoStop}%
\bibitem [{\citenamefont {Zhou}\ and\ \citenamefont {Ng}(2013)}]{Zhou2013}%
  \BibitemOpen
  \bibfield  {author} {\bibinfo {author} {\bibfnamefont {Yi}~\bibnamefont
  {Zhou}}\ and\ \bibinfo {author} {\bibfnamefont {Tai-Kai}\ \bibnamefont
  {Ng}},\ }\bibfield  {title} {\enquote {\bibinfo {title} {{Spin Liquid States
  in the Vicinity of a Metal-insulator Transition}},}\ }\href {\doibase
  10.1103/PhysRevB.88.165130} {\bibfield  {journal} {\bibinfo  {journal} {Phys.
  Rev. B}\ }\textbf {\bibinfo {volume} {88}},\ \bibinfo {pages} {165130}
  (\bibinfo {year} {2013})},\ \Eprint {http://arxiv.org/abs/1302.0157}
  {arXiv:1302.0157} \BibitemShut {NoStop}%
\bibitem [{\citenamefont {Reiter}(1994)}]{Reiter1994}%
  \BibitemOpen
  \bibfield  {author} {\bibinfo {author} {\bibfnamefont {George~F.}\
  \bibnamefont {Reiter}},\ }\bibfield  {title} {\enquote {\bibinfo {title}
  {{Self-consistent wave function for magnetic polarons in the t - J model}},}\
  }\href {\doibase 10.1103/PhysRevB.49.1536} {\bibfield  {journal} {\bibinfo
  {journal} {Phys. Rev. B}\ }\textbf {\bibinfo {volume} {49}},\ \bibinfo
  {pages} {1536--1539} (\bibinfo {year} {1994})}\BibitemShut {NoStop}%
\bibitem [{\citenamefont {Zhu}\ \emph {et~al.}(2014)\citenamefont {Zhu},
  \citenamefont {Jiang}, \citenamefont {Sheng},\ and\ \citenamefont
  {Weng}}]{Zhu2014}%
  \BibitemOpen
  \bibfield  {author} {\bibinfo {author} {\bibfnamefont {Zheng}\ \bibnamefont
  {Zhu}}, \bibinfo {author} {\bibfnamefont {Hong~Chen}\ \bibnamefont {Jiang}},
  \bibinfo {author} {\bibfnamefont {D.~N.}\ \bibnamefont {Sheng}}, \ and\
  \bibinfo {author} {\bibfnamefont {Zheng~Yu}\ \bibnamefont {Weng}},\
  }\bibfield  {title} {\enquote {\bibinfo {title} {{Nature of Strong Hole
  Pairing in Doped Mott Antiferromagnets}},}\ }\href {\doibase
  10.1038/srep05419} {\bibfield  {journal} {\bibinfo  {journal} {Sci. Rep.}\
  }\textbf {\bibinfo {volume} {4}},\ \bibinfo {pages} {5419} (\bibinfo {year}
  {2014})},\ \Eprint {http://arxiv.org/abs/1312.6893} {arXiv:1312.6893}
  \BibitemShut {NoStop}%
\bibitem [{\citenamefont {Weng}(2011)}]{Weng2011a}%
  \BibitemOpen
  \bibfield  {author} {\bibinfo {author} {\bibfnamefont {Zheng-Yu}\
  \bibnamefont {Weng}},\ }\bibfield  {title} {\enquote {\bibinfo {title}
  {{Superconducting Ground State of a Doped Mott Insulator}},}\ }\href
  {\doibase 10.1088/1367-2630/13/10/103039} {\bibfield  {journal} {\bibinfo
  {journal} {New J. Phys.}\ }\textbf {\bibinfo {volume} {13}},\ \bibinfo
  {pages} {103039} (\bibinfo {year} {2011})},\ \Eprint
  {http://arxiv.org/abs/1105.3027} {arXiv:1105.3027} \BibitemShut {NoStop}%
\bibitem [{\citenamefont {Wang}\ \emph {et~al.}(2015)\citenamefont {Wang},
  \citenamefont {Zhu}, \citenamefont {Qi},\ and\ \citenamefont
  {Weng}}]{Wang2015}%
  \BibitemOpen
  \bibfield  {author} {\bibinfo {author} {\bibfnamefont {Qing-Rui}\
  \bibnamefont {Wang}}, \bibinfo {author} {\bibfnamefont {Zheng}\ \bibnamefont
  {Zhu}}, \bibinfo {author} {\bibfnamefont {Yang}\ \bibnamefont {Qi}}, \ and\
  \bibinfo {author} {\bibfnamefont {Zheng-Yu}\ \bibnamefont {Weng}},\
  }\bibfield  {title} {\enquote {\bibinfo {title} {{Variational Wave Function
  for an Anisotropic Single-hole-doped $t$-$J$ Ladder}},}\ }\href@noop {} {\
  (\bibinfo {year} {2015})},\ \Eprint {http://arxiv.org/abs/1509.01260}
  {arXiv:1509.01260} \BibitemShut {NoStop}%
\bibitem [{\citenamefont {Zhao}\ \emph {et~al.}(2022)\citenamefont {Zhao},
  \citenamefont {Chen}, \citenamefont {Zhang},\ and\ \citenamefont
  {Weng}}]{Zhao2022}%
  \BibitemOpen
  \bibfield  {author} {\bibinfo {author} {\bibfnamefont {Jing-Yu}\ \bibnamefont
  {Zhao}}, \bibinfo {author} {\bibfnamefont {Shuai~A.}\ \bibnamefont {Chen}},
  \bibinfo {author} {\bibfnamefont {Hao-Kai}\ \bibnamefont {Zhang}}, \ and\
  \bibinfo {author} {\bibfnamefont {Zheng-Yu}\ \bibnamefont {Weng}},\
  }\bibfield  {title} {\enquote {\bibinfo {title} {{Two-Hole Ground State:
  Dichotomy in Pairing Symmetry}},}\ }\href {\doibase
  10.1103/PhysRevX.12.011062} {\bibfield  {journal} {\bibinfo  {journal} {Phys.
  Rev. X}\ }\textbf {\bibinfo {volume} {12}},\ \bibinfo {pages} {011062}
  (\bibinfo {year} {2022})},\ \Eprint {http://arxiv.org/abs/2106.14898}
  {arXiv:2106.14898} \BibitemShut {NoStop}%
\bibitem [{\citenamefont {Zhu}\ \emph {et~al.}(2016)\citenamefont {Zhu},
  \citenamefont {Wang}, \citenamefont {Sheng},\ and\ \citenamefont
  {Weng}}]{Zhu2016}%
  \BibitemOpen
  \bibfield  {author} {\bibinfo {author} {\bibfnamefont {Zheng}\ \bibnamefont
  {Zhu}}, \bibinfo {author} {\bibfnamefont {Qing-Rui}\ \bibnamefont {Wang}},
  \bibinfo {author} {\bibfnamefont {D.~N.}\ \bibnamefont {Sheng}}, \ and\
  \bibinfo {author} {\bibfnamefont {Zheng-Yu}\ \bibnamefont {Weng}},\
  }\bibfield  {title} {\enquote {\bibinfo {title} {{Exact Sign Structure of the
  $t$-$J$ Chain and the Single Hole Ground State}},}\ }\href {\doibase
  10.1016/j.nuclphysb.2015.12.004} {\bibfield  {journal} {\bibinfo  {journal}
  {Nucl. Phys. B}\ }\textbf {\bibinfo {volume} {903}},\ \bibinfo {pages}
  {51--77} (\bibinfo {year} {2016})},\ \Eprint
  {http://arxiv.org/abs/1510.07634} {arXiv:1510.07634} \BibitemShut {NoStop}%
\bibitem [{\citenamefont {Chen}\ \emph {et~al.}(2018)\citenamefont {Chen},
  \citenamefont {Zhu},\ and\ \citenamefont {Weng}}]{Chen2018}%
  \BibitemOpen
  \bibfield  {author} {\bibinfo {author} {\bibfnamefont {Shuai}\ \bibnamefont
  {Chen}}, \bibinfo {author} {\bibfnamefont {Zheng}\ \bibnamefont {Zhu}}, \
  and\ \bibinfo {author} {\bibfnamefont {Zheng-Yu}\ \bibnamefont {Weng}},\
  }\bibfield  {title} {\enquote {\bibinfo {title} {{Two-hole Ground State
  Wavefunction: Non-BCS Pairing in a $t$-$J$ Two-leg Ladder}},}\ }\href
  {\doibase 10.1103/PhysRevB.98.245138} {\bibfield  {journal} {\bibinfo
  {journal} {Phys. Rev. B}\ }\textbf {\bibinfo {volume} {98}},\ \bibinfo
  {pages} {245138} (\bibinfo {year} {2018})},\ \Eprint
  {http://arxiv.org/abs/1808.06173} {arXiv:1808.06173} \BibitemShut {NoStop}%
\bibitem [{\citenamefont {Zhang}\ and\ \citenamefont {Weng}(2022)}]{Zhang2022}%
  \BibitemOpen
  \bibfield  {author} {\bibinfo {author} {\bibfnamefont {Jia-Xin}\ \bibnamefont
  {Zhang}}\ and\ \bibinfo {author} {\bibfnamefont {Zheng-Yu}\ \bibnamefont
  {Weng}},\ }\bibfield  {title} {\enquote {\bibinfo {title} {{Crossover from
  Fermi Arc to Full Fermi Surface}},}\ }\href@noop {} {\ ,\ \bibinfo {pages}
  {1--20} (\bibinfo {year} {2022})},\ \Eprint {http://arxiv.org/abs/2208.10519}
  {arXiv:2208.10519} \BibitemShut {NoStop}%
\bibitem [{\citenamefont {Liang}\ \emph {et~al.}(1988)\citenamefont {Liang},
  \citenamefont {Doucot},\ and\ \citenamefont {Anderson}}]{Liang1988}%
  \BibitemOpen
  \bibfield  {author} {\bibinfo {author} {\bibfnamefont {S.}~\bibnamefont
  {Liang}}, \bibinfo {author} {\bibfnamefont {B.}~\bibnamefont {Doucot}}, \
  and\ \bibinfo {author} {\bibfnamefont {P.~W.}\ \bibnamefont {Anderson}},\
  }\bibfield  {title} {\enquote {\bibinfo {title} {{Some New Variational
  Resonating-Valence-Bond-Type Wave Functions for the Spin-$\frac{1}{2}$
  Antiferromagnetic Heisenberg Model on a Square Lattice}},}\ }\href {\doibase
  10.1103/PhysRevLett.61.365} {\bibfield  {journal} {\bibinfo  {journal} {Phys.
  Rev. Lett.}\ }\textbf {\bibinfo {volume} {61}},\ \bibinfo {pages} {365--368}
  (\bibinfo {year} {1988})}\BibitemShut {NoStop}%
\bibitem [{\citenamefont {Sandvik}\ and\ \citenamefont
  {Evertz}(2010)}]{Sandvik2010}%
  \BibitemOpen
  \bibfield  {author} {\bibinfo {author} {\bibfnamefont {Anders~W.}\
  \bibnamefont {Sandvik}}\ and\ \bibinfo {author} {\bibfnamefont {Hans~Gerd}\
  \bibnamefont {Evertz}},\ }\bibfield  {title} {\enquote {\bibinfo {title}
  {{Loop Updates for Variational and Projector Quantum Monte Carlo Simulations
  in the Valence-Bond Basis}},}\ }\href {\doibase 10.1103/PhysRevB.82.024407}
  {\bibfield  {journal} {\bibinfo  {journal} {Phys. Rev. B}\ }\textbf {\bibinfo
  {volume} {82}},\ \bibinfo {pages} {024407} (\bibinfo {year} {2010})},\
  \Eprint {http://arxiv.org/abs/0807.0682} {arXiv:0807.0682} \BibitemShut
  {NoStop}%
\end{thebibliography}%

\end{document}